\pdfoutput=1
\documentclass[PAPER, atlasdraft=false, texlive=2023, UKenglish, cernpreprint, texmf, orcidlogo]{atlasdoc}
\usepackage[subcaption]{atlaspackage}
\usepackage{atlasbiblatex}

\usepackage[hepprocess=true,jetetmiss=true]{atlasphysics}

\usepackage{physics}
\usepackage{verbatim}
\usepackage{tabularx}
\usepackage{multirow}

\graphicspath{{logos/}{figures/}}

\usepackage{ANA-EXOT-2020-04-PAPER-defs}

\AtlasTitle{Search for dark matter produced in association with a dark Higgs boson decaying into \(W^{+}W^{-}\) in the one-lepton final state at \(\sqrt{s} = 13\,\)TeV using 139\(\,\)fb\(^{-1}\) of \(pp\) collisions recorded with the ATLAS detector}

\AtlasAbstract{%
Several extensions of the Standard Model predict the production of dark matter particles at the LHC.
A search for dark matter particles produced in association with a dark Higgs boson decaying into \(W^{+}W^{-}\) in the $\ell^\pm\nu q \bar q'$ final states with \(\ell=e,\mu\) is presented.
This analysis uses 139~fb\(^{-1}\) of \(pp\) collisions recorded by the ATLAS detector at a centre-of-mass energy of 13~TeV.
The \(W^\pm \to q\bar q'\) decays are reconstructed from pairs of calorimeter-measured jets or from track-assisted reclustered jets, a technique aimed at resolving the dense topology from a pair of boosted quarks using jets in the calorimeter and tracking information.
The observed data are found to agree with Standard Model predictions.
Scenarios with dark Higgs boson masses ranging between 140 and 390~GeV are excluded.
}

\AtlasRefCode{EXOT-2020-04}

\AtlasNote{EXOT-2020-04}

\PreprintIdNumber{CERN-EP-2022-147}

\HepDataRecord{132484}
\AtlasJournalRef{JHEP 07 (2023) 116}
\AtlasDOI{10.1007/JHEP07(2023)116}

\hypersetup{pdftitle={ATLAS document},pdfauthor={The ATLAS Collaboration}}

\begin{document}

\maketitle

\section{Introduction}
\label{sec:intro}

% The next lines are included from the .//sections/introduction.tex input file
An overwhelming body of astrophysical evidence~\cite{Bertone:2010zza,Feng:2010gw,Porter:2011nv,Bertone:2017adx} primarily from galactic rotation velocity measurements, from the gravitational lensing effect and from the spectral analysis of the cosmic microwave background in the context of the Big Bang nucleosynthesis strongly supports the existence of dark matter (DM).
The Standard Model of particle physics (SM) does not provide a viable DM candidate.
The particle nature of DM is one of the major questions in physics.

Several extensions of the SM postulate a DM candidate \wimp that is a stable, electrically neutral, and weakly interacting massive particle (WIMP)~\cite{Bertone:2017adx}.
Such WIMPs can potentially be produced in high-energy collisions at the CERN Large Hadron Collider (LHC).
Their production at the LHC would be characterized by a striking signature with an imbalance in the measured transverse momentum\footnote{\AtlasCoordFootnote} from the WIMPs escaping detection, denoted as missing transverse momentum~\ptmiss with magnitude \(\met\equiv|\ptmiss|\).
Hence, a wide class of DM models probed at the LHC postulate processes where one or more SM particles \(X\) are produced recoiling against DM particles, resulting in an `\(X+\met\)' signature. Searches at the LHC have considered \(X\) to be a hadronic jet~\cite{EXOT-2018-06,CMS-EXO-20-004}, top or bottom quarks~\cite{EXOT-2018-43,SUSY-2018-34,SUSY-2018-08,SUSY-2018-07,CMS-EXO-16-049,CMS-EXO-18-010}, a photon~\cite{EXOT-2018-63,CMS-EXO-16-039}, a \W or \Z boson~\cite{EXOT-2016-23,HIGG-2016-28,CMS-EXO-19-003,CMS-EXO-16-048}, or a Higgs boson~\cite{EXOT-2018-46,CMS-EXO-18-011,CMS-EXO-16-055}.

Recently, exploration of a new \(X+\met\) DM signature began at the LHC, where \(X\) is a hypothetical particle that is  produced in proton--proton (\(pp\)) collisions and decays into a vector-boson pair \(VV=W^+W^-\) or \(ZZ\).
It was first probed by the ATLAS Collaboration in the fully hadronic decay channel \(VV\to q\bar q'\! q''\!\bar q'''\), where the invariant mass of the diboson system, \(m_{VV}\), is above 160~\GeV~\cite{EXOT-2018-40}.

This paper presents an exploration of the hitherto uncharted semileptonic decay channel \(W^+W^-\to \ell^\pm\nu \qpqbar\).
In the signal region~(SR), events are required to have large \MET from DM particles escaping detection and the neutrino from the leptonically decaying \(W\) boson, an energetic electron or a muon, and a hadronically decaying \(W\) boson.
Candidate \(\Wtoqq\) decays are identified as a pair of small-radius jets with an invariant mass consistent with the \(W\) boson mass \(m_W\), or, if sufficiently boosted, i.e.\ with a sufficiently high \pt, as a single large-radius jet with a similar invariant mass requirement.
The latter category dominates the sensitivity, due to significantly smaller backgrounds. Large-radius jets are identified using the track-assisted reclustered (TAR) jet reconstruction technique~\cite{ATL-PHYS-PUB-2018-012} that was first employed in Ref.~\cite{EXOT-2018-40}.
The background is dominated by vector-boson production in association with jets, referred to as \vjets in the following.
The analysis employs a control region (CR), defined by requiring a large separation in polar angle between the hadronic \(W\) candidate and the charged lepton, to improve the modelling of the \wjets background by determining its yield from data.
Another CR requiring two or more \(b\)-quark jets is used to constrain the contribution of the \ttbar background and to  improve its modelling in the SR.

\begin{figure}[t]
\centering
\begin{subfigure}{0.31\textwidth}
\includegraphics[width = \textwidth]{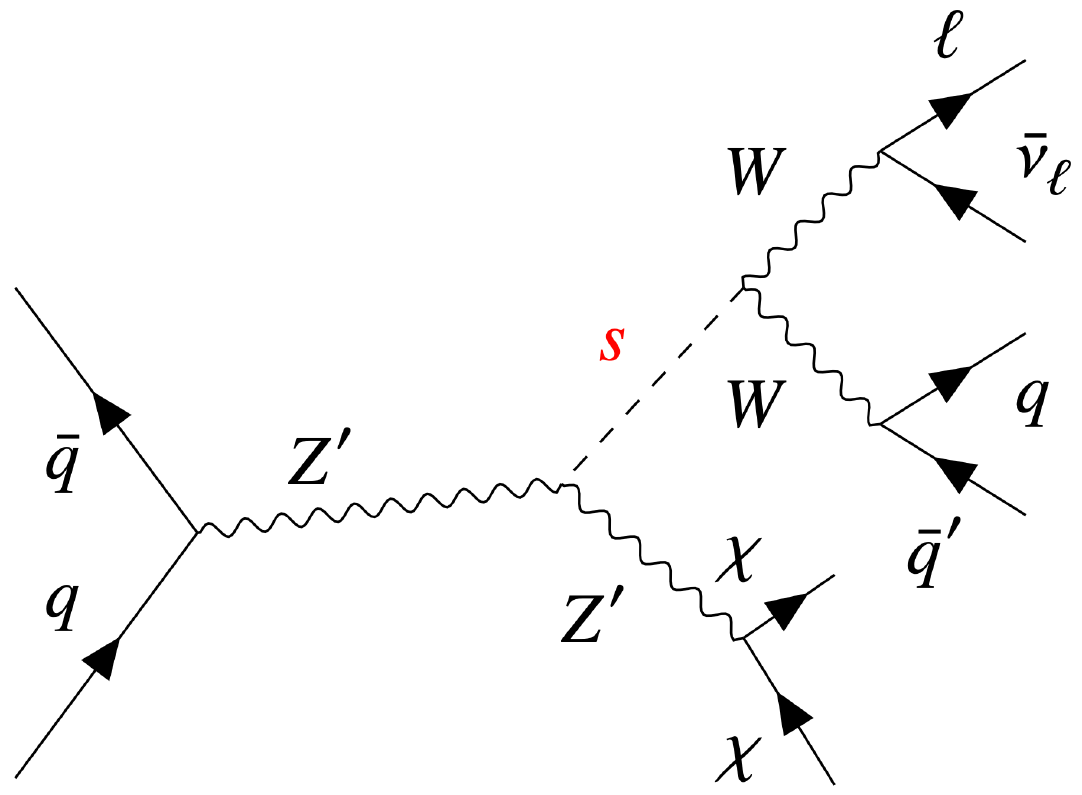}
\caption{}
\end{subfigure}
\quad
\begin{subfigure}{0.35\textwidth}
\includegraphics[width = \textwidth]{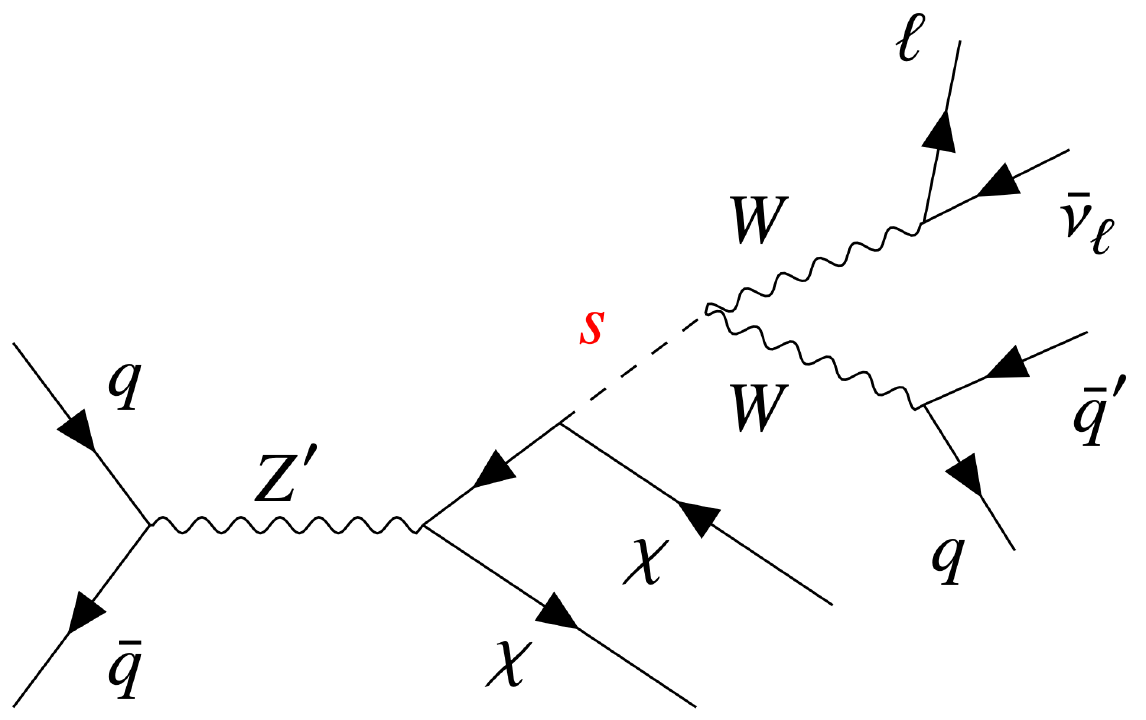}
\caption{}
\end{subfigure}
\quad
\begin{subfigure}{0.24\textwidth}
\includegraphics[width = \textwidth]{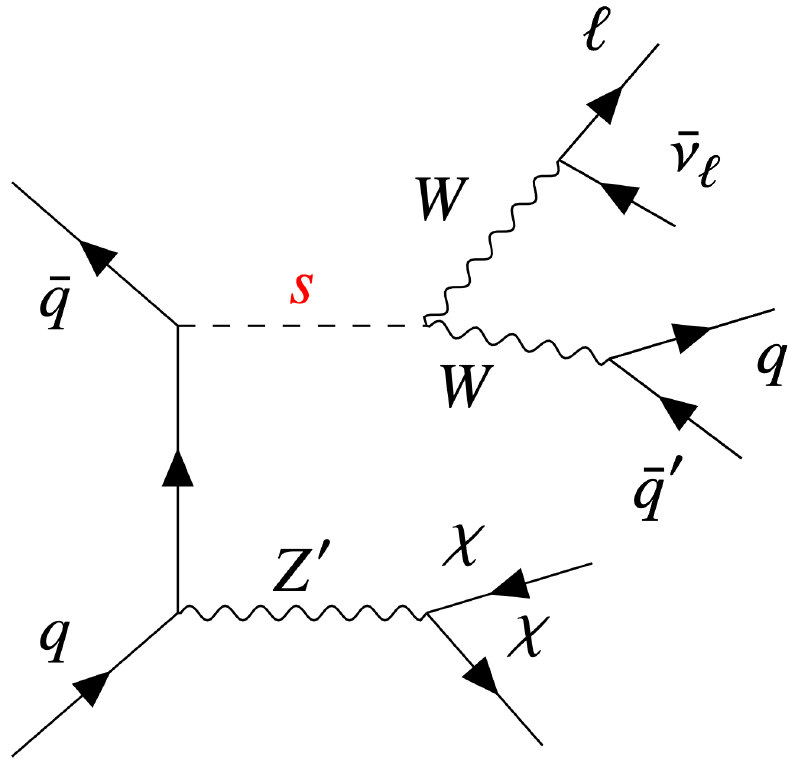}
\caption{}
\end{subfigure}
\caption{
Representative Born-level Feynman diagrams for resonant \(q\bar q \to Z' \to s\chi\chi\) production where the dark Higgs boson decays into semileptonic final states, \(s\to W^-(\ell^-\bar\nu) W^+(\qpqbar)\). Diagram (a) typically dominates for high \mZp, while diagram (c) contributes most for \(\mZp\lesssim 2\mDM + \ms\). Diagram (b) makes a sizeable contribution throughout the parameter space considered. Charge-conjugated processes are implied.
}
\label{fig:feyn}
\end{figure}

The discovery of a new boson~\cite{HIGG-2012-27,CMS-HIG-12-028} consistent with the expectations for the SM Higgs boson confirmed the Brout--Englert--Higgs (BEH) mechanism for electroweak symmetry breaking~\cite{Englert:1964et,Higgs:1964ia,Higgs:1964pj,Guralnik:1964eu,Higgs:1966ev,Kibble:1967sv} that is ultimately responsible for the generation of masses for SM particles.
The experimental confirmation of the BEH mechanism motivates a similar mechanism for mass generation in the dark sector that contains the DM particle~\cite{Kahlhoefer:2015bea}.
In this theoretical paradigm, a fermionic DM particle obtains its mass through Yukawa interactions with a dark Higgs boson, \s~\cite{Duerr:2017uap}.
Another motivation for this theoretical scenario is the provision of a new DM annihilation channel into SM particles involving the dark Higgs boson when it, rather than the WIMP DM particle, is the lightest state in the dark sector.
This feature addresses the  stringent experimental constraints from the observed DM relic density~\cite{Aghanim:2018eyx}.
The dark Higgs boson can be identified with the aforementioned resonance decaying into a pair of massive vector bosons, \(s\to VV\), and produced in association with DM, which strongly motivates exploring the \(s(\to VV)+\met\) signature presented in this paper.
This signature is complementary to the $\text{jet}+\met$ signature, which is always present for DM production at the LHC irrespective of the underlying model, since WIMPs result in $\met$ and additional jet(s) can be produced from initial-state radiation.
A two-mediator-based DM model~\cite{Duerr:2016tmh} is used for the optimization and interpretation of the search presented in this paper.
This model features a new \Up gauge symmetry, which yields an additional massive spin-1 vector boson \Zp through the BEH mechanism involving a new complex Higgs field and producing the new physical dark Higgs boson \s.
The characteristic model parameters are the mass \mDM of the Majorana DM particle, which is a singlet under all SM symmetries, the \Zp mass \mZp, the dark Higgs mass \ms, the \Zp couplings \gq to quarks, the \Zp couplings \gchi to DM particles, and the mixing angle \(\theta\) between the SM and dark Higgs bosons~\cite{Duerr:2017uap}, which is set to a small value~\cite{EXOT-2018-40}.
Representative Born-level Feynman diagrams for the signal processes targeted in this search are shown in Figure~\ref{fig:feyn}.
The \sDM signal is produced through the \(q\bar q \to Z' \to s\chi\chi\) process, which requires an off-shell intermediate state such as a \Zp or \(\chi\).
The dark Higgs boson interacts with SM particles only via mixing with the SM Higgs boson, resulting in the same decay branching fractions as a SM Higgs boson with a different mass as long as the \stoDM process is kinematically forbidden.
The \sWW and \sZZ processes become important for \(\ms \gtrsim 160~\GeV\) and \(\ms \gtrsim 180~\GeV\), respectively~\cite{Heinemeyer:2013tqa}.
The proposed dark Higgs two-mediator DM model framework shares similarities with previously explored spin-1 simplified DM models~\cite{Buchmueller:2013dya,Harris:2014hga,Buckley:2014fba,Abercrombie:2015wmb,Abdallah:2015ter,Albert:2017onk}, where \s is the only new additional particle and \wimp is a Majorana fermion rather than a Dirac fermion.
All choices of common model parameters are made in alignment with Ref.~\cite{Albert:2017onk} to facilitate comparisons between experiments and between search channels.
Other searches in \(pp\) collisions for a spin-1 mediator that is identified with the \Zp boson in the two-mediator-based DM model provide complementary sensitivity~\cite{EXOT-2017-32}.
The strongest limits come from dijet searches, which exclude the full \mZp range investigated in this paper with the chosen parameters \cite{Argyropoulos:2021sav}.
Nevertheless, the exploration of the \(s(\to VV)+\met\) signature is still strongly motivated, as the aforementioned complementarity relies on the assumptions about the underlying simplified model which may not be realised in a complete model, and on the choice of couplings like \gchi and \gq.

The paper is organized as follows. The ATLAS detector and the analysed data set together with simulations are discussed in Sections~\ref{sec:detector} and~\ref{sec:data-mc}. The reconstruction of events is covered in Section~\ref{sec:object-reconstruction}, which is followed by a description of the track-assisted reclustering algorithm in Section~\ref{sec:tar-jets}. The data analysis procedures are summarized in Section~\ref{sec:analysis} and the corresponding systematic uncertainties are estimated in Section~\ref{sec:systematics}. The results are presented in Section~\ref{sec:results} and conclusions are drawn in Section~\ref{sec:conclusion}.

% End of text imported from the .//sections/introduction.tex input file

%
\section{ATLAS detector}
\label{sec:detector}

% The next lines are included from the .//sections/atlas-detector.tex input file
The ATLAS detector~\cite{PERF-2007-01} at the LHC covers nearly the entire solid angle around the collision point.
It consists of an inner tracking detector (ID) surrounded by a thin superconducting solenoid, electromagnetic and hadronic calorimeters, and a muon spectrometer incorporating three  superconducting toroidal magnets.

The ID system is immersed in a \SI{2}{\tesla} axial magnetic field and provides charged-particle tracking in the range \(|\eta|{\,<\,}2.5\).
The high-granularity silicon pixel detector covers the vertex region and typically provides four measurements per track~\cite{ATLAS-TDR-2010-19,PIX-2018-001}.
It is followed by the silicon microstrip tracker, which usually provides eight measurements per track for central \(|\eta|\).
These silicon detectors are complemented by the transition radiation tracker, which enables radially extended track reconstruction up to \(|\eta|{\,=\,}2.0\).
This detector also provides electron identification information based on the fraction of hits (typically 30 in total) above a higher energy-deposit threshold corresponding to transition radiation.

The calorimeter system covers the pseudorapidity range \(|\eta|{\,<\,}4.9\).
Within the region \(|\eta|{\,<\,} 3.2\), electromagnetic calorimetry is provided by barrel and endcap high-granularity lead/liquid-argon (LAr) calorimeters,
with an additional thin LAr presampler covering \(|\eta|{\,<\,}1.8\) to correct for energy loss in material upstream of the calorimeters.
Hadronic calorimetry is provided by the steel/scintillator-tile calorimeter, segmented into three barrel structures within \(|\eta|{\,<\,}1.7\), and two copper/LAr hadronic endcap calorimeters.
The solid angle coverage is completed with forward copper/LAr and tungsten/LAr calorimeter modules optimized for electromagnetic and hadronic measurements respectively.

The muon spectrometer comprises separate trigger and high-precision tracking chambers measuring the deflection of muons in a magnetic field generated by the superconducting air-core toroids.
The field integral of the toroids ranges between \num{2.0} and \SI{6.0}{\tesla\metre} across most of the detector.
A set of precision chambers covers the region \(|\eta|{\,<\,}2.7\) with three layers of monitored drift tubes, complemented by cathode-strip chambers in the forward region, where the background is highest.
The muon trigger system covers the range \(|\eta|{\,<\,}2.4\) with resistive-plate chambers in the barrel, and thin-gap chambers in the endcap regions.

Events are selected to be recorded by the first-level (L1) trigger system implemented in custom hardware, followed by selections made by algorithms implemented in software in the high-level trigger (HLT)~\cite{TRIG-2016-01}.
The L1 trigger accepts events from the \SI{40}{\MHz} bunch crossings at a rate below \SI{100}{\kHz}, which the HLT reduces in order to record events to disk at about \SI{1}{\kHz}.

An extensive software suite~\cite{ATL-SOFT-PUB-2021-001} is used in the reconstruction and analysis of real and simulated data, in detector operations, and in the trigger and data acquisition systems of the experiment.

% End of text imported from the .//sections/atlas-detector.tex input file

%
\section{Data and simulated events}
\label{sec:data-mc}

% The next lines are included from the .//sections/data-mc.tex input file
The data used in this analysis are  proton--proton collisions provided by the LHC during 2015--2018 and recorded by the ATLAS detector.
The data were taken at a centre-of-mass energy of \(\sqrt{s} = \SI{13}{\TeV}\) with a minimum separation of \SI{25}{ns} between consecutive crossings of proton bunches from the two beams.
The data set corresponds to an integrated luminosity of \SI{139}{\ifb}, determined by using the LUCID-2 detector \cite{LUCID2} for the primary luminosity measurement. The corresponding uncertainty is \SI{1.7}{\%} \cite{ATLAS-CONF-2019-021}.
The data were collected using a missing transverse momentum trigger in combination with single-muon triggers \cite{TRIG-2019-04,TRIG-2019-01,TRIG-2018-01}.
The missing transverse momentum trigger is based on \(\boldsymbol{p}_{\text{T}}^{\text{miss, trigger}}\) as computed from calorimeter information.
Depending on the data-taking period, the missing transverse momentum trigger was one of a set with thresholds ranging from \(E_{\text{T}}^{\text{miss, trigger}} = 70~\GeV\) to \(E_{\text{T}}^{\text{miss, trigger}} = 110~\GeV\).
In the absence of muons, this set of triggers was measured to be fully efficient for events with \(\met > \SI{200}{\GeV}\), as reconstructed using offline algorithms.
Events with muons can escape the \met triggers.
This is because muons tend to deposit negligible amounts of energy in the calorimeters and hence contribute to \(\boldsymbol{p}_{\text{T}}^{\text{miss, trigger}}\) that is reconstructed using exclusively calorimeter information.
This can result in a lower  \(E_{\text{T}}^{\text{miss, trigger}}\) for signal events in kinematic configurations where the muon counterbalances the \pt from the neutrino and the WIMPs.
To compensate for the decreased efficiency of the \met trigger, which is particularly pronounced for high-\pt muons, a combination of single-muon triggers was added.
Some of these with an isolation criterion applied a low \pt threshold ranging from \SI{20}{\GeV} to \SI{26}{\GeV} in different data-taking periods, while others without an isolation criterion applied a \pt threshold ranging from \SI{50}{\GeV} to \SI{60}{\GeV}.
In events selected by a muon trigger, an offline reconstructed muon is required to match the trigger-level muon.
This combination of a missing transverse momentum trigger and single-muon triggers was measured to be fully efficient for the examined final states after the selections described in Section~\ref{sec:preselection}.

Monte Carlo (MC) simulations were used to model the expected kinematic behaviour of SM background processes as well as the investigated signal.
They are described in the following, starting with the SM background processes relevant to this analysis.

The \wjets production process was simulated with the \SHERPA[2.2.10] generator~\cite{Bothmann:2019yzt} using next-to-leading-order (NLO) matrix elements for up to two partons, and leading-order (LO) matrix elements for up to four partons, calculated with the Comix~\cite{Gleisberg:2008fv} and \OPENLOOPS~\cite{Buccioni:2019sur,Cascioli:2011va,Denner:2016kdg} libraries.
They were matched with the \SHERPA parton shower~\cite{Schumann:2007mg} using the \MEPSatNLO prescription~\cite{Hoeche:2011fd,Hoeche:2012yf,Catani:2001cc,Hoeche:2009rj} and the set of tuned parameters developed by the \SHERPA authors.
The samples were normalized to a next-to-next-to-leading-order (NNLO) prediction~\cite{Anastasiou:2003ds}.
The \zjets process was generated using an identical set-up, except that \SHERPA[2.2.11] was used and LO matrix elements were calculated for up to five partons.

Samples of diboson final states (\(VV\)) were simulated with the \SHERPA[2.2.1] or 2.2.2 generator depending on the process, including off-shell effects and Higgs boson contributions where appropriate.
Fully leptonic final states and semileptonic final states, where one boson decays leptonically and the other hadronically, were generated using matrix elements at NLO accuracy in QCD for up to one additional parton and at LO accuracy for up to three additional parton emissions.
Samples for the loop-induced processes \(gg \to VV\) were generated using LO-accurate matrix elements for up to one additional parton emission.
The matrix element calculations were matched and merged with the \SHERPA parton shower based on Catani--Seymour dipole factorization~\cite{Gleisberg:2008fv,Schumann:2007mg} using the MEPS@NLO prescription.
The virtual QCD corrections were provided by the \OPENLOOPS library.

The electroweak \(VVjj\) production processes, which include vector-boson scattering, vector-boson fusion Higgs production, and triboson processes with this final state when taking interference into account, as well as triboson (\(VVV\)) processes with other final states,
were simulated with the \SHERPA[2.2.2] generator using factorized gauge-boson decays.
To account for interference, the two sets of triboson processes are treated together and labelled `triboson' in the following.
Matrix elements, accurate to NLO for the inclusive process and to LO for up to two additional parton emissions (\(VVV\)) or accurate to LO (\(VVjj\)), were matched and merged with the \SHERPA parton shower based on Catani--Seymour dipole factorization using the \MEPSatNLO prescription.
The virtual QCD corrections for matrix elements at NLO accuracy were provided by the \OPENLOOPS library.

All \SHERPA weak-boson samples were generated using the \NNPDF[3.0nnlo] set of  parton distribution functions (PDFs)~\cite{Ball:2014uwa}, along with the dedicated set of tuned parton-shower parameters developed by the \SHERPA authors.
For all weak-boson processes, uncertainties from missing higher orders were evaluated~\cite{Bothmann:2016nao} using seven variations of the QCD factorization and renormalization scales in the matrix elements by factors of \(0.5\) and \(2\), excluding variations in opposite directions.
Uncertainties associated with the PDF set were evaluated according to the PDF4LHC recommendations~\cite{Butterworth:2015oua}:
uncertainties in the nominal PDF set were calculated using 100 replica variations.
The effect of the uncertainty in the strong coupling constant \(\alphas\) was assessed by variations of \(\pm 0.001\).
Uncertainties arise from the scale choices for the matching of the matrix elements (CKKW)~\cite{Lonnblad:2011xx} and the resummation calculations to the \SHERPA parton-shower algorithm. These uncertainties are evaluated by changing the nominal CKKW scale of \SI{20}{\GeV} to either \SI{15}{\GeV} or \SI{30}{\GeV} and by varying the resummation scale by a factor of \(0.5\) or \(2\). The same relative uncertainties were assumed for \(VVV\) and \(VVjj\) backgrounds per analysis region.
For \zjets, all uncertainties were extrapolated from \wjets in order to avoid large effects from statistical fluctuations.

The production of \ttbar events was modelled using the \POWHEGBOX[v2]~\cite{Frixione:2007nw,Nason:2004rx,Frixione:2007vw,Alioli:2010xd} generator at NLO with the \NNPDF[3.0nlo] PDF set and the \hdamp parameter
set to 1.5\,\mtop~\cite{ATL-PHYS-PUB-2016-020}.
The \hdamp parameter is a resummation damping factor and one of the parameters that controls the matching of \POWHEG matrix elements to the parton shower and thus effectively regulates the high-\pt radiation against which the \ttbar system recoils.
The events were interfaced to \PYTHIA[8.230]~\cite{Sjostrand:2014zea} to model the parton shower, hadronization, and underlying event, with parameter values set according to the A14 tune~\cite{ATL-PHYS-PUB-2014-021} and using the \NNPDF[2.3lo] set of PDFs~\cite{Ball:2012cx}.
The decays of bottom and charm hadrons in all MC samples involving top quarks were performed by \EVTGEN[1.6.0]~\cite{Lange:2001uf}.

The production of top quarks in association with \(W\) bosons (\(tW\)) was modelled by the \POWHEGBOX[v2]~\cite{Re:2010bp,Nason:2004rx,Frixione:2007vw,Alioli:2010xd} generator at NLO in QCD, using the five-flavour scheme and the \NNPDF[3.0nlo] set of PDFs.
The diagram removal scheme~\cite{Frixione:2008yi} was used to remove interference and overlap with \ttbar production.
The related uncertainty was estimated by comparing these events with an alternative sample generated using the diagram subtraction scheme~\cite{Frixione:2008yi,ATL-PHYS-PUB-2016-020}.
The events were interfaced to \PYTHIA[8.230], which used the A14 tune and the \NNPDF[2.3lo] set of PDFs.

Single-top \(t\)-channel production was modelled by the \POWHEGBOX[v2]~\cite{Frederix:2012dh,Nason:2004rx,Frixione:2007vw,Alioli:2010xd} generator at NLO in QCD, using the four-flavour scheme and the corresponding \NNPDF[3.0nlo] set of PDFs.
The events were interfaced with \PYTHIA[8.230], which used the A14 tune and the \NNPDF[2.3lo] set of  PDFs. Single-top \(s\)-channel production was studied and found to be negligible.

For all processes involving top quarks, the uncertainty due to initial-state radiation was estimated by simultaneously varying the \hdamp parameter and the renormalization and factorization scales \muR and \muF, and choosing the Var3c up/down variants of the A14 tune as described in Ref.~\cite{ATL-PHYS-PUB-2017-007}.
The impact of final-state radiation was evaluated by raising or lowering the renormalization scale for emissions from the parton shower by a factor two.
To evaluate the PDF uncertainties for the nominal PDF, the 100 variations for \NNPDF[3.0nlo] were taken into account; for \ttbar the effect of a \(\pm 0.001\) change in \alphas was also evaluated.
The uncertainty due to the parton-shower and hadronization model was evaluated by comparing the nominal samples of events with samples where events generated by \POWHEGBOX[v2]~\cite{Re:2010bp,Nason:2004rx,Frixione:2007vw,Alioli:2010xd} were interfaced to \HERWIG[7.04]~\cite{Bahr:2008pv,Bellm:2015jjp}, using the H7UE set of tuned parameters~\cite{Bellm:2015jjp} and the \MMHT[lo] PDF set~\cite{Harland-Lang:2014zoa}.
To assess the uncertainty in the matching of NLO matrix elements to the parton shower, the \POWHEGBOX samples where compared with samples of events generated with \MGNLO[2.6.0] interfaced with \PYTHIA[8.230].
The \MGNLO calculation used the \NNPDF[3.0nlo] set of PDFs, and \PYTHIA[8] used the A14 tune and the \NNPDF[2.3lo] set of PDFs.

Simulated event samples for the \(p p \rightarrow Z' \rightarrow s \chi \chi \rightarrow W^{+} W^{-} \chi \chi \rightarrow \qpqbar \ell^{\pm} \nu \chi \chi\) process were generated at LO in QCD with up to one additional parton in the event, using \MGNLO[2.8.1]~\cite{2014alwall} with the \NNPDF[3.0nlo] PDF set.
Other processes arising from the dark Higgs model could also contribute to the analysis regions of this search but were not considered. These include the \(s \rightarrow hh\) decay channel, which contributes a small fraction of the signal for $m_s>250~\GeV$.
The parton-showering process was simulated with \PYTHIA[8.244] using the A14 tune and the \NNPDF[2.3lo] set of PDFs.
Samples were generated in the (\(m_{Z'}\), \(m_s\)) plane for \(0.3 \leq m_{Z'} \leq \SI{3.3}{\TeV}\) and for \(115 \leq m_s \leq \SI{385}{\GeV}\).
Other dark Higgs model parameter values were chosen as \(m_\chi = \SI{200}{\GeV}\) to avoid $s\to\chi\chi$ decays in the $m_s$ range considered, \(g_\chi = 1.0\), \(g_q = 0.25\)~\cite{Abercrombie:2015wmb,Albert:2017onk}, and \(\sin{\theta} = 0.01\)~\cite{EXOT-2018-40}.
Uncertainties due to QCD factorization and renormalization were estimated using seven variations of the scales by factors  of \(0.5\) and \(2\), avoiding variations in opposite directions.
Uncertainties from the PDF set were estimated using 100 replica variations.
The uncertainties due to the parton-shower and hadronization model were estimated by comparing the nominal samples with alternative samples from the same matrix element generator but interfaced with \HERWIG[7.2.1], using the H7UE set of tuned parameters and the \NNPDF[3.0nlo] PDF set.

The effect of multiple interactions in the same and neighbouring bunch crossings (\pileup) was modelled by overlaying the simulated hard-scattering event with inelastic \(pp\) events generated with \PYTHIA[8.186]~\cite{Sjostrand:2007gs} using the \NNPDF[2.3lo] set of PDFs and the A3 tune~\cite{ATL-PHYS-PUB-2016-017}.
The detector response for all MC samples was modelled with a full detector simulation~\cite{SOFT-2010-01} based on \GEANT~\cite{Agostinelli:2002hh}.

% End of text imported from the .//sections/data-mc.tex input file

%
\section{Event reconstruction}
\label{sec:object-reconstruction}

% The next lines are included from the .//sections/object-reconstruction.tex input file
%
Each event requires the presence of at least one \pp collision vertex that is reconstructed from at least two ID tracks with \(\pttrk > 0.5~\GeV\). The vertex with the highest \(\sum(\pttrk)^2\) in the event is designated the primary vertex (PV).
The ID tracks must have at least seven hits and satisfy \(\pt>0.5~\GeV\) and \(|\eta|<2.5\) requirements~\cite{ATL-PHYS-PUB-2015-006,ATL-PHYS-PUB-2015-051}. Their transverse and longitudinal impact parameters relative to the PV must satisfy \(|\dzero|<2~\mathrm{mm}\) and \(|\zzsth|<3~\mathrm{mm}\), respectively, where \(\theta\) is the polar angle of the track.

Charged leptons \(\ell=e,\mu\) are used to identify events with leptonic final states produced in decays of \(W\) bosons.
This includes \(\tau\)-lepton decays into \(\ell\). Hadronic \(\tau\)-lepton decays are not considered due to their small contribution to the overall sensitivity of the analysis.
Electrons~\cite{EGAM-2018-01} are reconstructed by matching clusters of energy deposited in the calorimeter to an ID track. Candidate electrons are identified using a likelihood-based method, and must satisfy the `Medium' requirement and be matched to the PV of the event.
Furthermore, they must fall within \(|\eta|<2.47\), have \(\pt > 7~\)\GeV, and be isolated from  additional activity reconstructed in the calorimeter and the
tracker, following the `Loose' isolation requirement.
To veto additional electrons in the event, the identification requirement is relaxed to satisfy the `Loose' working point and have a hit in the innermost layer of the pixel detector, while the isolation requirement remains unchanged.
Muons~\cite{MUON-2018-03} are reconstructed by matching a track or track segment detected in the muon spectrometer, depending on the \pt and \(|\eta|\) range, to an ID track that is matched to the PV.
Muon candidates must satisfy `Medium' requirements, have \(\pt>7~\GeV\) and fall within \(|\eta|<2.5\).
Furthermore, they must be isolated from  additional activity reconstructed in the tracker within a cone of \pt-dependent angular size $\Delta R$ around the muon, following the `FixedCutTightTrackOnly' isolation requirement.
To veto additional muons, the identification requirement is relaxed to satisfy the `Loose' working point and the muon is required to fall within \(|\eta|<2.7\), which corresponds to the acceptance region of the muon spectrometer, while the isolation requirement is dropped.
The selections for leptons are summarized in Table~\ref{tab:leptons}.
Electrons and muons which satisfy only the relaxed definition are labelled as `veto leptons'.

\begin{table}
\centering
\caption{
Overview of the electron and muon selection criteria.
\label{tab:leptons}
}
\begin{tabular}{lllll}
\toprule
\textbf{Feature} & \textbf{Electron} & \textbf{Veto electron} & \textbf{Muon} & \textbf{Veto muon} \\
\midrule
\midrule
Pseudorapidity range & \(\abs{\eta}<2.47\) & \(\abs{\eta}<2.47\) & \(\abs{\eta} < 2.5\) & \(\abs{\eta} < 2.7\) \\
Transverse momentum & \(\pt > 7~\GeV\) & \(\pt > 7~\GeV\) & \(\pt > 7~\GeV\) & \(\pt > 7~\GeV\) \\
\midrule
Identification & Medium & Loose & Medium & Loose \\
Isolation & Loose & Loose & FixedCutTightTrackOnly & - \\
\bottomrule
\end{tabular}
\end{table}

To avoid reconstruction ambiguities, if a muon candidate and an electron candidate share a track, the muon candidate is ignored from further analysis if it is calorimeter-tagged, and the electron candidate otherwise.
Calorimeter-tagged muons are identified as an ID track that can be matched only to an energy deposit in the calorimeter compatible with a minimum-ionizing particle.
Since muons of this type have no muon spectrometer information, they have lower purity than other muon types~\cite{MUON-2018-03}.

Jets of hadrons are used to identify the hadronically decaying \(W\) bosons and other hadronic activity in the event.
Particle-flow (PFlow) jets are constructed from charged constituents associated with the PV and neutral constituents~\cite{PERF-2015-09}  using the \antikt algorithm~\cite{Cacciari:2008gp,Fastjet} with a radius parameter of \(R=0.4\).
The energy of PFlow jets is calibrated to the particle scale using a sequence of corrections, including \pileup subtraction and in situ calibration~\cite{JETM-2018-05}, and the energy resolution in MC simulations is calibrated to match that found in data.
PFlow jets are required to have \(\pt{\,>\,}20\,\GeV\) and fall within the acceptance region of the ID, i.e.\ \(|\eta|<2.5\).
The jet vertex tagger (JVT) discriminant is applied to reject jets with \(\pt<60\)~\GeV\ and \(|\eta|<2.4\) originating from pile-up interactions through the use of tracking and vertexing information~\cite{PERF-2014-03}.
PFlow jets closer than \(\DeltaR = 0.2\) to an electron are rejected. A similar requirement is applied to PFlow jets with fewer than three tracks in the vicinity of a muon. Electrons or muons at an angular distance of less than \(\Delta R=\min(0.4, 0.04 + 10\,\GeV/\pt^\ell)\) from a jet are rejected, as they are likely to originate from hadron decays within jets. Here, $\Delta R\propto 1/\pt^\ell$ accounts for tighter collimation of electromagnetic showers from electrons and final-state radiation from electrons and muons with increasing \pt.

The decay products of boosted \(W\) bosons with \(\pt\gtrsim150~\GeV\) are collimated, which makes the matching of the energy deposits to the individual quarks from the \(\Wtoqq\) decay challenging.
This is addressed through jets that are formed from the energy in three-dimensional clusters of calorimeter cells using the \antikt algorithm  with a small radius parameter of \(R=0.2\).
The energy of the input calorimeter cells is corrected for dead material, out-of-cluster losses for pions, and calorimeter response to hadronic energy clusters, which are identified using their topology and energy density. These corrections are implemented with a local cell signal weighting (LCW) method~\cite{PERF-2014-07}.
Corrections for \pileup and the energy scale and resolution---following methodologies similar to those in \Refn{\cite{PERF-2016-04}}---are subsequently applied to \smallR jets.
The \smallR jets must have \(\pt{\,>\,}20\,\GeV\) and fall within the acceptance region of the ID, i.e.\ \(|\eta|<2.5\).
The reconstruction of highly boosted, hadronically decaying \(W\) bosons is discussed in Section~\ref{sec:tar-jets}.

In order to suppress contributions from background processes that involve top quarks, which decay almost exclusively to \(b\)-quarks, a multivariate deep-learning algorithm, DL1r, is used to identify (\(b\)-tag) jets containing \(b\)-hadrons. The \(b\)-tagging is performed using PFlow jets with a working point corresponding to an efficiency of 77\%~\cite{PERF-2016-05} for jets containing \(b\)-hadrons. Events containing \(b\)-tagged jets are rejected.

The \ptmiss vector is computed as the negative vector sum of the transverse momenta of the veto electrons, veto muons, PFlow jet candidates, and ID tracks not associated with the aforementioned objects~\cite{PERF-2016-07}.
Hadronically decaying \(\tau\)-leptons and photons are not explicitly considered in the \ptmiss calculation.
In addition, an object-based \met significance \(\mathcal{S}\) is used to reject multijet background events. The \(\mathcal{S}\) observable is computed from the expected resolutions for all the objects used in the \met calculation~\cite{ATLAS-CONF-2018-038}.

% End of text imported from the .//sections/object-reconstruction.tex input file

%
\section{Track-assisted reclustering}
\label{sec:tar-jets}

% The next lines are included from the .//sections/tar-jets.tex input file

%
%
%

With increasing momenta of the dark Higgs boson, its decay products and ultimately the products of the \(\Wtoqq\) decay become more boosted and hence more collimated.
The track-assisted reclustering (TAR) algorithm~\cite{ATL-PHYS-PUB-2018-012} is employed to reconstruct the topological substructure of boosted \(\Wtoqq\) decays.
This algorithm improves the resolution of jet substructure observables by considering both tracking and calorimeter information, combined with the flexibility of jet reclustering.
The TAR jets are formed from \smallR jets reclustered into larger jets with \(R=1\) using trimming parameters optimized for the ATLAS experiment~\cite{ATL-PHYS-PUB-2015-033}.
The trimming procedure was shown to improve the mass resolution of \(W\) decay jets despite the \pileup corrections to the input \smallR jets.
Any \smallR jets overlapping with an electron within \(\Delta R=0.2\) are not considered in the reclustering process.
The mass and other substructure observables of TAR jets are reconstructed using ID tracks.
For this, ID tracks are first matched to the \smallR jets that constitute the \(R=1\) jets using a two-step procedure.
First, ID tracks are matched to \smallR jets using ghost association~\cite{Cacciari:2008gn}. Second, any remaining tracks closer than \(\Delta R=0.3\) to \smallR jets are matched to the closest \smallR jet in \(\Delta R\).
Tracks associated with electrons or muons are not considered in the matching procedure.
Subsequently, the \pt of tracks matched to a given \smallR jet are rescaled such that their sum equals the \pt of that jet:

\begin{equation*}
\pt^\text{track,\,new} = \pt^\text{track,\,old} \times \frac{ p_{\text{T}}^\text{\smallR} }{ \sum_{i}p_{\text{T},i}^\text{track,\,old}}\,,
\label{eq:track-assist}
\end{equation*}

where the sum in \(i\) runs over all tracks matched to a given \smallR jet.
The jet mass and other jet substructure observables of a reclustered \(R=1\) jet are then reconstructed from these rescaled tracks.
This procedure compensates for the neutral jet components missed by the tracker and improves the reconstruction of jet substructure observables, which is limited by the angular resolution of the detector~\cite{ATL-PHYS-PUB-2018-012}.
Furthermore, the TAR algorithm provides a straightforward method to calibrate and assign
uncertainties to jet substructure observables by propagating calibrations and uncertainties from the well-defined
constituent \smallR jets and tracks, as is detailed in Section~\ref{sec:systematics}.

\begin{figure}
\centering
\includegraphics[width=0.3\textwidth]{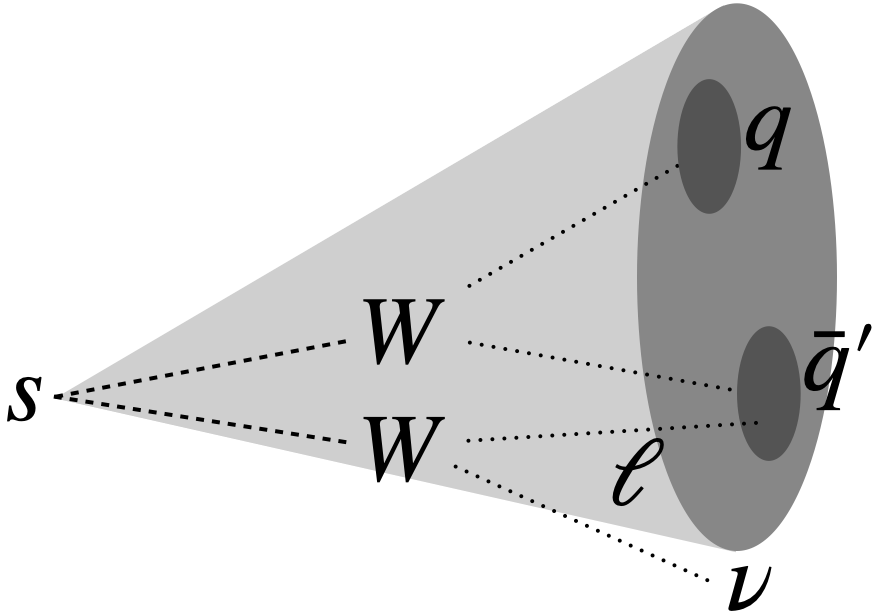}
\caption{
Sketch of the \(s\to WW\to \ell\nu \qpqbar\) decay where the dark Higgs boson is significantly boosted, reconstructed as a large-\(R\) jet in the detector. The collimation due to the boost of the dark Higgs decay products often leads to an overlap of the charged lepton \(\ell\) and the large-\(R\) jet from the \(\Wtoqq\) decay. %
}
\label{fig:sketch_overlap}
\end{figure}

As already mentioned above, TAR jets with \(R=1\) are used to reconstruct \(\Wtoqq\) candidate decays, leading to a two-prong jet signature.
When the final-state particles from the \(s\to WW\to \ell\nu \qpqbar\) decays are collimated due to a significant boost of the dark Higgs boson, this leads to a dense environment with the charged lepton often being very close to the hadronic \(W\) decay products, as schematically shown in \Fig{\ref{fig:sketch_overlap}}.
In this situation, additional tracks from charged leptons and calorimetric energy clusters from electrons tend to fall into the catchment area of the \(R=1\) jet used to reconstruct the \(\Wtoqq\) decay. For `standard' large-\(R\) jets~\cite{JETM-2018-02}, this was shown to distort the kinematic properties and substructure of \(W\) boson candidate jets, resulting in a significant reduction of their identification efficiency. This effect is particularly pronounced in the \(s\to WW\to e\nu_e \qpqbar\) decay channel. With the TAR algorithm, this challenge can be met by appropriately preselecting the input tracks and \smallR jets, rejecting those associated with leptons, as described above. This ensures that only hadronic objects are considered in the reconstruction of the \(\Wtoqq\) candidates using TAR jets. Hence, the typical two-prong structure with a reconstructed mass near the \(W\) boson mass is preserved. The small radius parameter of  \(R=0.2\) used for the reconstruction of \antikt subjets forming TAR jets prevents significant numbers of hadronic components from being removed because of their proximity to electrons.

The criteria used to identify \(W\) candidates in this search are the TAR jet mass and the ratio of energy correlation functions \Dtwo~\cite{2014larkoski}, which is a measure for judging whether the topological substructure of a jet is consistent with a two-prong decay.
The identification criteria are chosen to optimize the \(\Wtoqq\) identification efficiency while effectively rejecting fake \(W\) candidates from QCD jets present in the dominant \wjets background processes. This is further described in \Sect{\ref{sec:signalregions}}.

% End of text imported from the .//sections/tar-jets.tex input file

%
\section{Analysis}
\label{sec:analysis}

% The next lines are included from the .//sections/analysis.tex input file
%
%
%
%

As a reminder, the target signature of this analysis, \(s\to WW\to \ell\nu \qpqbar\), is characterized by a single charged lepton \(\ell\), missing transverse momentum, and a hadronically decaying \(W\) boson candidate \Whad.
The SR is split into two categories based on the reconstruction method used for \Whad: a \emph{merged} category, where the hadronically decaying \(W\) boson candidate is reconstructed as a single \(R=1\) TAR jet, and a \emph{resolved} category, where the candidate consists of two \(R=0.4\) jets.
The merged reconstruction method is characterized by its excellent background rejection, but also a relatively low signal efficiency.
This motivates the addition of the resolved category to accept signal events that fail to meet the strict criteria for the merged category but still contribute to the overall sensitivity.
The orthogonality of the merged and resolved categories is ensured by prioritizing the more sensitive merged category: events fulfilling its selection criteria are not considered for the resolved category.

Dedicated control regions (CR) are constructed to improve the modelling of the \wjets and \ttbar background processes by determining their normalizations from data. These CRs are referred to as CRW and CRTT, respectively.
Each CR is split into a merged and a resolved category using the same approach as in the SR to closely match its kinematics.
The final discriminant to separate signal from background in the SR is \msmin, which represents the minimum possible mass of the dark Higgs boson that is compatible with the signal process given the kinematic properties of the measured reconstructed objects; this observable is introduced in detail in \Sect{\ref{sec:minms}}.
No \msmin information is used in the CRs that are used to constrain the overall yield of the \wjets and \ttbar processes.
The individual data analysis steps are described in greater detail in the following.

\subsection{Preselection}
\label{sec:preselection}
Events that are found to contain any jets with properties consistent with beam-induced backgrounds, cosmic-ray showers, or noisy calorimeter cells are rejected~\cite{DAPR-2018-01}.
All events are required to have exactly one lepton and no additional veto leptons.
A missing transverse momentum of \(\met > \SI{200}{\GeV}\) is required, with a significance of \(\metsig > 5\).
At least one TAR jet or at least two \(R=0.4\) jets are required.
Since a majority of SM background events with this signature originate from events where a single neutrino from a \(W \rightarrow \lnu\) decay is the only source of \ptmiss,  a leptonic transverse mass of \(\mt > \SI{150}{GeV}\) is required.
The leptonic transverse mass is defined as
\begin{equation*}
\label{eq:mtlep}
\mt = \sqrt{2 \ptlep \met \left( 1- \cos{\left( \phi_\ell - \phi_{\ptmiss} \right)} \right) }\,,
\end{equation*}
where the subscript \(\ell\) indicates kinematic observables of the charged lepton.
Stricter criteria are applied depending on the analysis region, as described in the following sections.

\subsection{Signal regions}
\label{sec:signalregions}
The selection requirements for the SR are summarized in \Tab{\ref{tab:mergedRegions}} for the merged category and in \Tab{\ref{tab:resolvedRegions}} for the resolved category.
In the following, a qualitative description of the used observables and the motivation for their use is given.

\begin{table}[htbp]
\centering
\caption{Selection criteria for the \merged category.
The two CRs are analysed inclusively, i.e.\ not considering any \msmin information.
In this category the $W$ boson candidate \Wcand is the \pt-leading TAR jet.}
\label{tab:mergedRegions}
\begin{tabular}{c| c c c}
\toprule
\textbf{Requirement}  &  \textbf{SR} &  \textbf{CRW} & \textbf{CRTT}  \\
\midrule
\midrule
Trigger & \multicolumn{3}{c}{\met or single muon} \\
\Nlep & \multicolumn{3}{c}{\(=1\)} \\
\mt[GeV] &  \multicolumn{3}{c}{\(>220\)} \\
\met[GeV] & \multicolumn{3}{c}{\(>200\)} \\
\midrule
\Nbjets &  0 & 0 & \(\geq 2\) \\
\NTAR  &  \multicolumn{3}{c}{\(\geq 1\)} \\
\Wcandm[GeV] &  \multicolumn{3}{c}{\([68,89]\)} \\
\metsig  &  \(>16\) &  {\(>12\)} &  {\(>12\)} \\
\dRWl &  \(<1.2\) & \(>1.8\) & \(<1.2\) \\
\Dtwo  &  \multicolumn{3}{c}{\(<1.1\)} \\
\midrule
\multirow{2}{*}{\msmin binning [GeV]} &  \([125,165,190,\) & \multirow{2}{*}{incl.} & \multirow{2}{*}{incl.} \\
&  \(225,375]\) \\
\bottomrule
\end{tabular}
\end{table}

\begin{table}[htbp]
\centering
\caption{Selection criteria for the \resolved category.
The two CRs are analysed inclusively, i.e.\ not considering any \msmin information.
In this category the $W$ boson candidate \Wcand is a combination of two $R = 0.4$ PFlow jets}
\label{tab:resolvedRegions}
\begin{tabular}{c| c c c}
\toprule
\textbf{Requirement}  &  \textbf{SR} &  \textbf{CRW} & \textbf{CRTT}  \\
\midrule
\midrule
Orthogonality & \multicolumn{3}{c}{Fails merged category selections} \\
\midrule
Trigger & \multicolumn{3}{c}{\met or single muon} \\
\Nlep & \multicolumn{3}{c}{\(=1\)} \\
\mt[GeV] &  \multicolumn{3}{c}{\(>200\)} \\
\met[GeV] & \multicolumn{3}{c}{\(>250\)} \\
\midrule
\Nbjets &  0 & 0 & \(\geq 2\) \\
\Njets  &  \multicolumn{3}{c}{\(\geq 2\)} \\
\Wcandm[GeV] &  \multicolumn{3}{c}{\([65,95]\)} \\
\metsig  &  \multicolumn{3}{c}{\(>16\)} \\
\drWl &  \(<1.4\) & \(>1.4\) & \(<1.4\) \\
\Wcandpt[GeV] &  \multicolumn{3}{c}{\(>150\)} \\
\midrule
\multirow{2}{*}{\msmin binning [GeV]} & \([125, 175, 225,\) & \multirow{2}{*}{incl.} & \multirow{2}{*}{incl.} \\
& \(275, 325, 375]\) \\
\bottomrule
\end{tabular}
\end{table}

\begin{figure}[htbp]
\centering
\begin{subfigure}{0.48\textwidth}
\includegraphics[width = \textwidth]{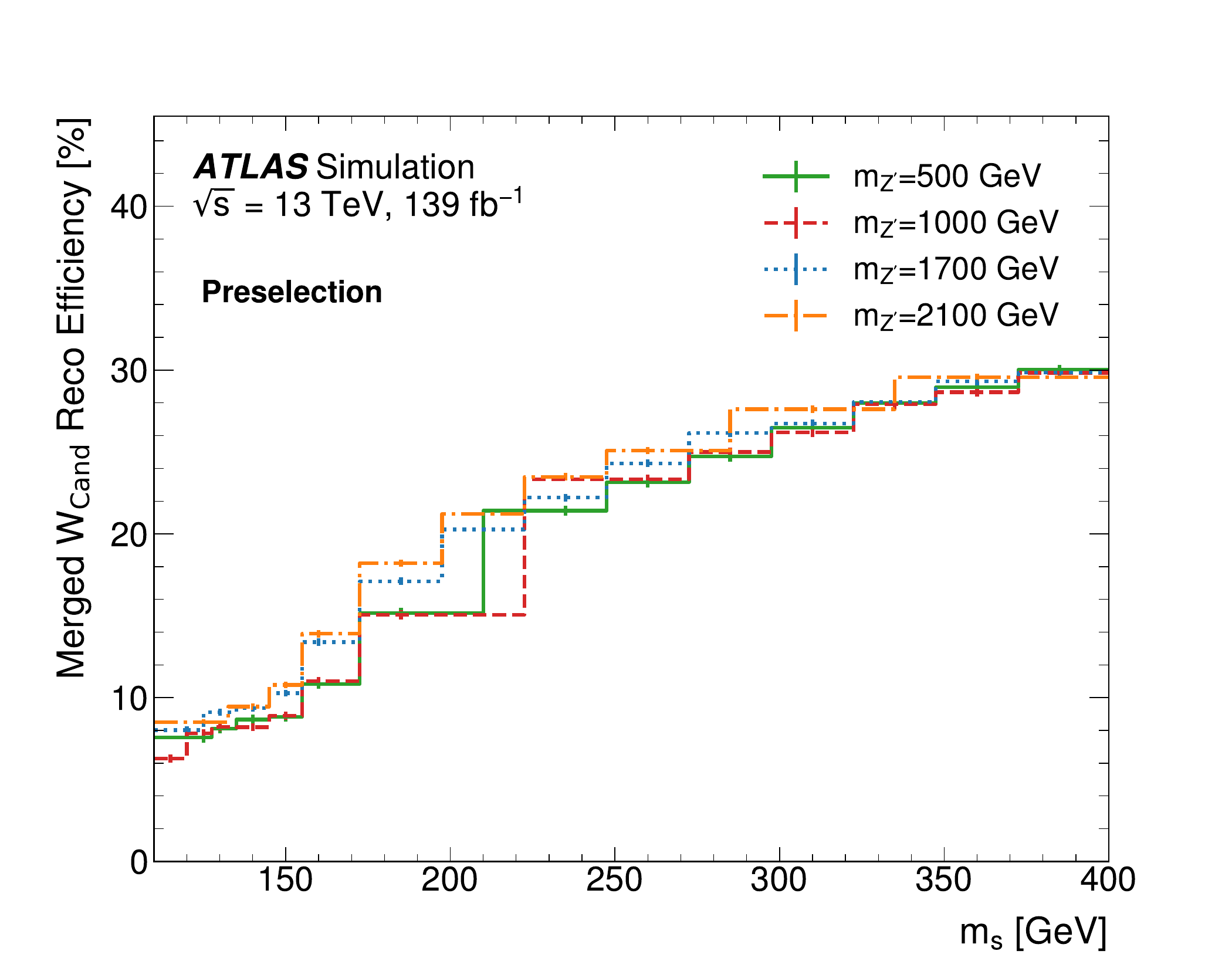}
\caption{Merged category}
\end{subfigure}
\quad
\begin{subfigure}{0.48\textwidth}
\includegraphics[width = \textwidth]{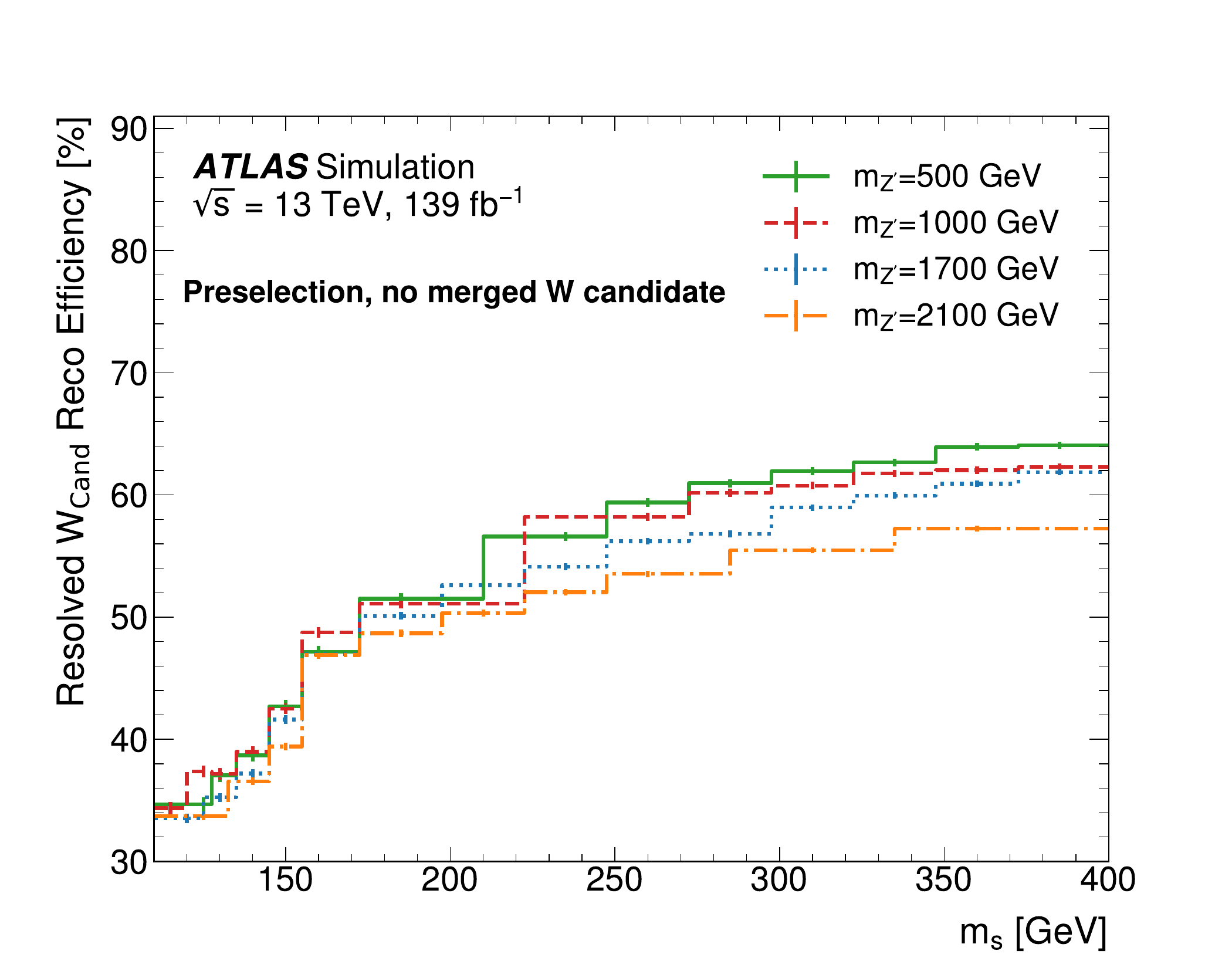}
\caption{Resolved category}
\end{subfigure}
\caption{
Probability of finding (a)~at least one TAR jet, where the \(\pt\)-leading TAR jet passes the \Wcandm and \Dtwo requirements in \Tab{\ref{tab:mergedRegions}},  as a function of \ms.
The probability is determined in a sample of signal events passing the preselections from \Sect{\ref{sec:preselection}}.
Probability of finding (b)~a \Whad candidate reconstructed as a pair of $R=0.4$ PFlow jets following requirements in \Tab{\ref{tab:resolvedRegions}},  as a function of \ms.
The probability is determined in a sample of signal events passing the preselections from \Sect{\ref{sec:preselection}} but not all of the requirements of the merged category.
The statistical uncertainties due to finite MC samples are indicated as vertical error bars.
}
\label{fig:TARefficiency}
\end{figure}

In the merged category, the \pt-leading TAR jet is used to reconstruct the hadronically decaying \(W\) boson candidate \Wcand.
The invariant mass of the TAR jet, \Wcandm, is required to be close to the \(W\) boson mass \(m_W\), which is taken to be 80.4~\GeV~\cite{Zyla:2020zbs}.
Additionally, the topological jet substructure observable \Dtwo~\cite{2014larkoski} is employed to select for a two-prong structure consistent with a \(\Wtoqq\) decay.
Figure~\ref{fig:TARefficiency}(a) shows the efficiency of this reconstruction method, defined as the fraction of events with at least one TAR jet, where the \(\pt\)-leading TAR jet passes the \Wcandm and \Dtwo requirements as described in \Tab{\ref{tab:mergedRegions}}, in a sample of signal events passing the preselection requirements from \Sect{\ref{sec:preselection}}. This efficiency ranges between 5\% and 30\% as a function of \ms.
For the dominant \wjets background, the corresponding efficiency is $(1.35\pm0.38)\%$, where the uncertainties are statistical only.

In the resolved category, the \Whad candidate is reconstructed as a combination of two \(R=0.4\)  PFlow jets.
If several combinations of jets are possible, the pair whose invariant mass is closest to \(m_W\) is chosen.
Since the \(W\) boson is expected to be significantly boosted due to the recoil from the DM particles, a high transverse momentum \Wcandpt is required for the \Whad candidate. As in the merged category, the invariant mass \Wcandm of the \Whad candidate is required to be close to \(m_W\).
The efficiency of finding a  \Whad candidate fulfilling these criteria as summarized in \Tab{\ref{tab:resolvedRegions}} is shown in Figure~\ref{fig:TARefficiency}(b). This efficiency is determined in a sample of events passing the preselections from \Sect{\ref{sec:preselection}} but not all of the requirements of the merged category, and ranges between 55\% and 65\% as a function of \ms.
For the dominant \wjets background, the corresponding efficiency is $(31.3 \pm 2.7)\%$ considering only statistical uncertainties.

\begin{figure}[htbp]
\centering
\begin{subfigure}{0.48\textwidth}
\includegraphics[width = \textwidth]{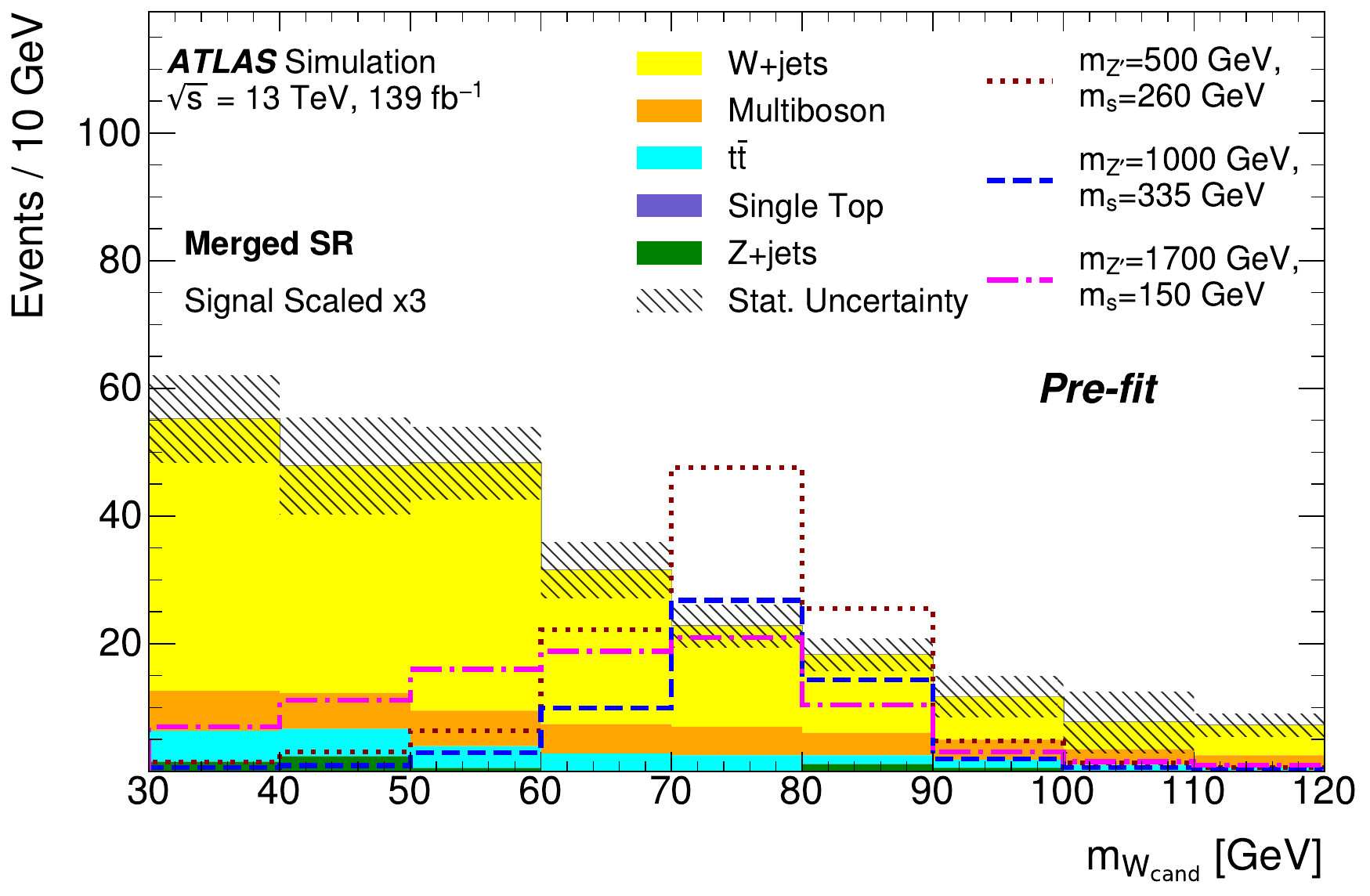}
\caption{Merged category}
\end{subfigure}
\quad
\begin{subfigure}{0.48\textwidth}
\includegraphics[width = \textwidth]{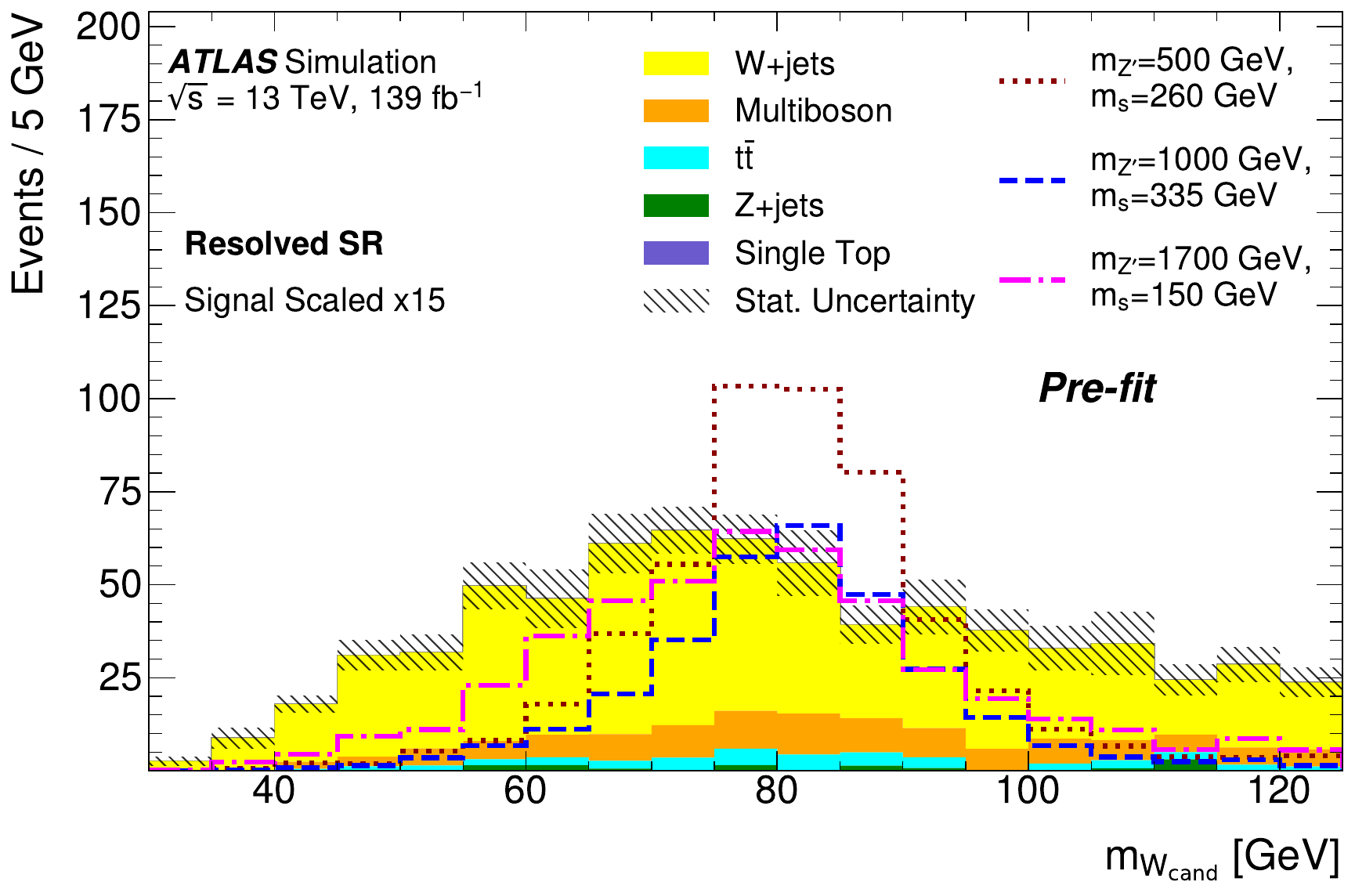}
\caption{Resolved category}
\end{subfigure}
\caption{Distributions of \Wcandm in the  signal region for the merged~(a) and resolved~(b) category, before any fit~(`Pre-fit').
The contributions from all SM backgrounds are shown as a histogram stack.
The `multiboson'  category includes contributions from \(VV\), \(VVV\), and electroweak \(VVjj\) production.
The hatched bands represent the total MC statistical uncertainty of the SM expectation.
The expected signal from a representative set of dark Higgs models with \(\gq=0.25\), \(\gchi=1\), \(\mDM=200~\GeV\), and \(\sin\theta=0.01\) scaled for presentation purposes is also shown for reference.
All selection criteria for the corresponding category (see \Tabs{\ref{tab:mergedRegions}}{\ref{tab:resolvedRegions}}) are applied, except the one for the displayed observable.
\label{fig:Wcand-mass}
}
\end{figure}

The mass distributions of the respective \Whad candidates in the merged and resolved categories are shown in \Fig{\ref{fig:Wcand-mass}}.
The \Wcandm window in the merged category is tighter than in the resolved category to maximize the sensitivity.

\begin{figure}[htbp]
\centering
\begin{subfigure}{0.48\textwidth}
\includegraphics[width = \textwidth]{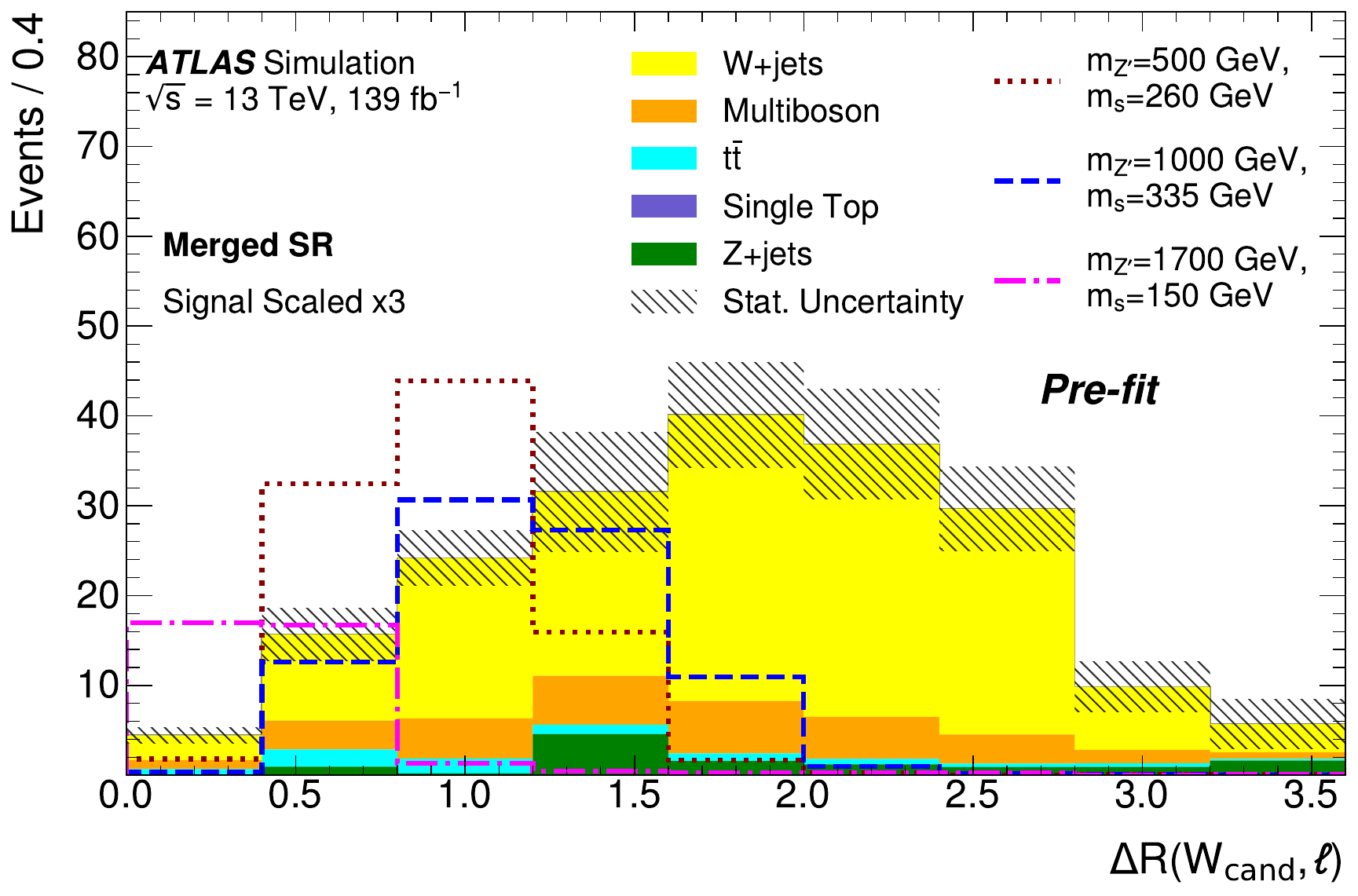}
\caption{Merged category}
\end{subfigure}
\quad
\begin{subfigure}{0.48\textwidth}
\includegraphics[width = \textwidth]{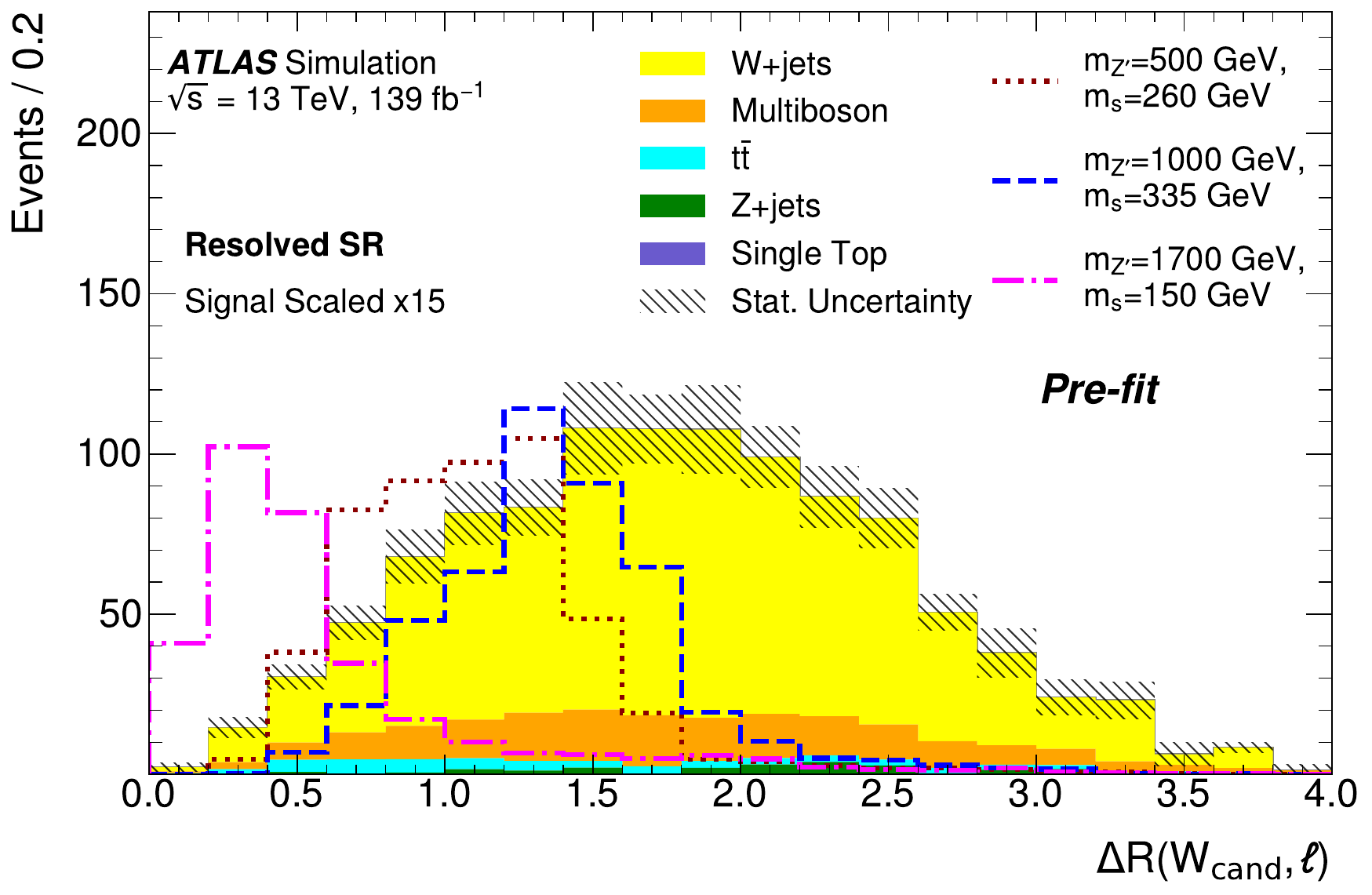}
\caption{Resolved category}
\end{subfigure}
\caption{
Distributions of \dRWl  in the signal region for the merged~(a) and resolved~(b) category, before any fit~(`Pre-fit').
The contributions from all SM backgrounds are shown as a histogram stack.
The `multiboson' category includes contributions from \(VV\), \(VVV\), and electroweak \(VVjj\) production.
The hatched bands represent the total MC statistical uncertainty of the SM expectation.
The expected signal from a representative set of dark Higgs models with \(\gq=0.25\), \(\gchi=1\), \(\mDM=200~\GeV\), and \(\sin\theta=0.01\) scaled for presentation purposes is also shown for reference.
All selection criteria for the corresponding category (see \Tabs{\ref{tab:mergedRegions}}{\ref{tab:resolvedRegions}}) are applied, except the one for the displayed observable.
\label{fig:DRWl}
}
\end{figure}

An important kinematic feature of signal events that distinguishes them from SM background events is that the momenta of the two \(W\) bosons are almost collinear because they recoil against an energetic pair of DM particles.
This leads to a small angular separation between the charged lepton and the hadronic \(W\) candidate.
By contrast, that angular separation is expected to be large for \wjets or diboson processes or other backgrounds where the hadronic \(W\) candidate does not originate from a real \(W\) boson.
Hence, an upper bound is placed on \dRWl, as shown in \Tabs{\ref{tab:mergedRegions}}{\ref{tab:resolvedRegions}} for the merged and resolved categories, respectively.
The distributions of this observable in both categories are shown in \Fig{\ref{fig:DRWl}}.

Events containing any \bjets are rejected in order to suppress the background contribution from \ttbar production.
The exact requirements on the observables described here and in \Sect{\ref{sec:preselection}} were optimized by maximizing the expected Asimov discovery significance~\cite{ATL-PHYS-PUB-2020-025}, which is calculated as

\begin{equation*}
Z_\text{sign} \equiv \left[ 2(s+b)\left(
\ln\left[ \frac{(s+b)(b+\sigma_b^2)}{b^2 + (s+b)\sigma_b^2} \right]
- \frac{b^2}{\sigma_b^2}\ln\left[ 1 + \frac{\sigma_b^2 s}{b(b+\sigma_b^2)} \right]
\right) \right]^\frac{1}{2}\,,
\label{eq:asimov}
\end{equation*}

where \(s\) and \(b\) are the expected total number of signal events and background events, respectively, and \(\sigma_b\) is the statistical uncertainty of the expected total number of background events.
The optimization was performed within the merged and resolved categories separately.
The typical acceptance $\times$ efficiency of this analysis for the studied dark Higgs model is about 1\% in both the merged and resolved categories.

\subsection{Dark Higgs reconstruction strategy}
\label{sec:minms}
\begin{figure}[htbp]
\centering
\begin{subfigure}{0.48\textwidth}
\includegraphics[width = \textwidth]{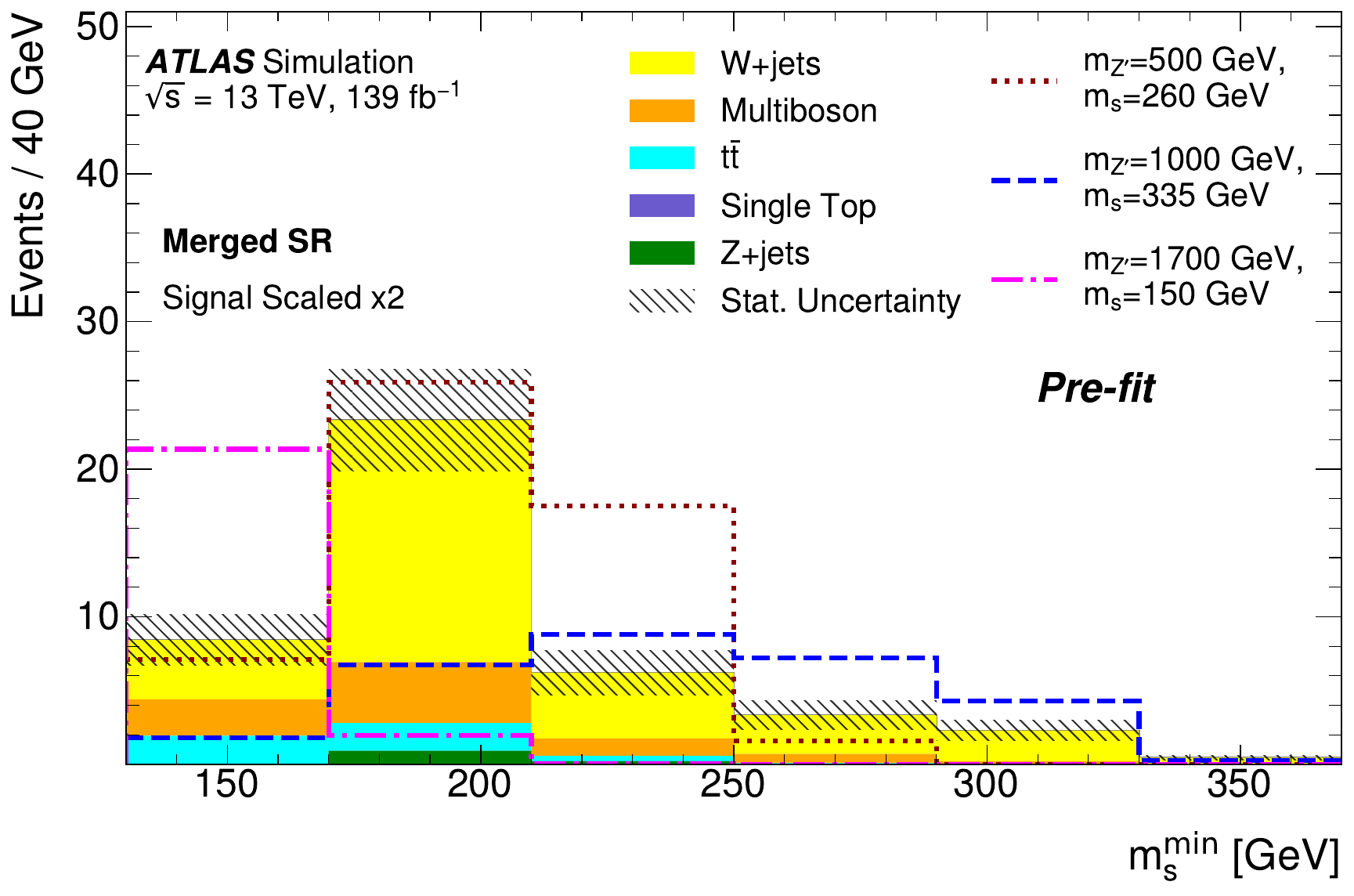}
\caption{Merged category}
\end{subfigure}
\quad
\begin{subfigure}{0.48\textwidth}
\includegraphics[width = \textwidth]{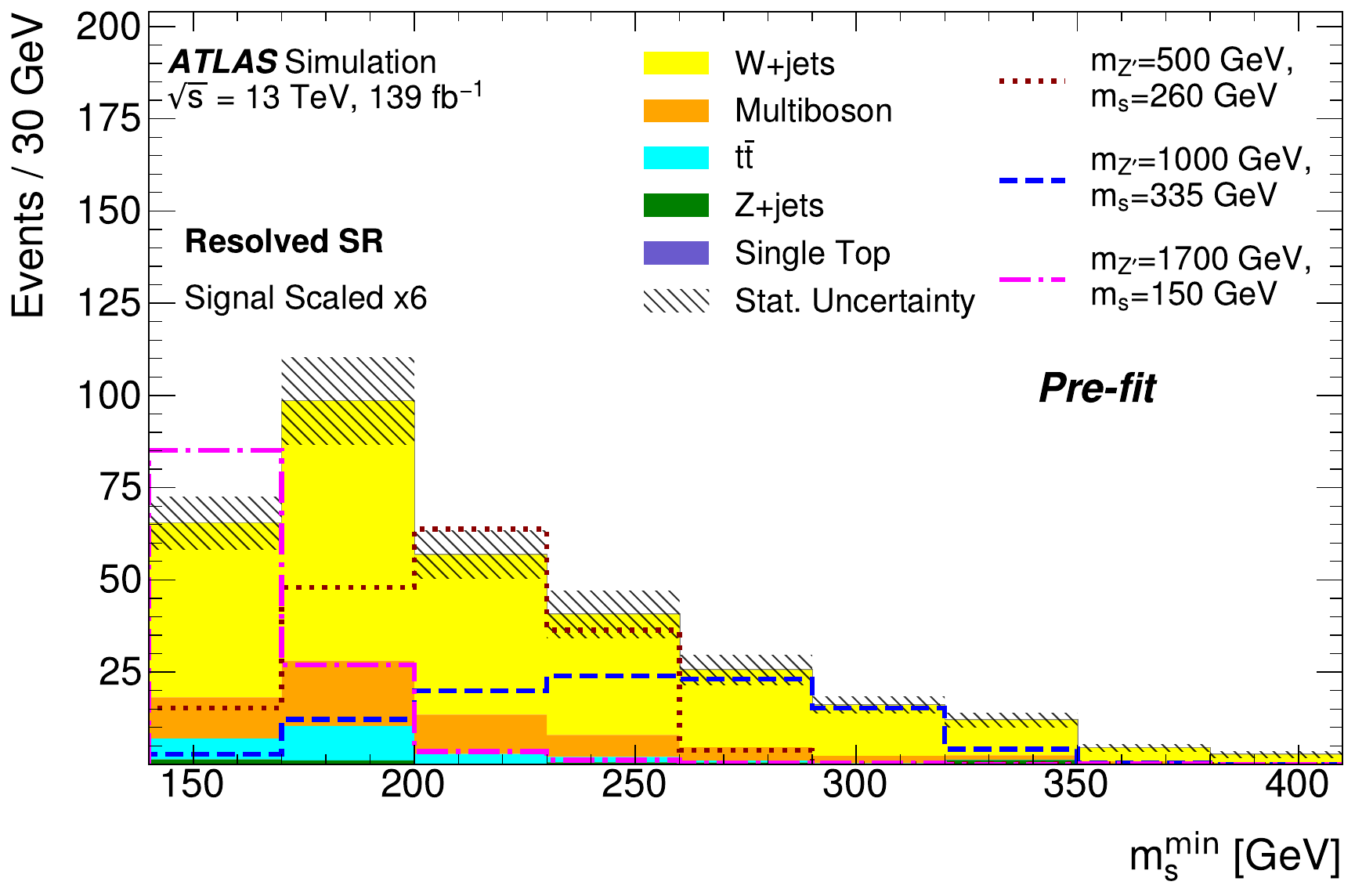}
\caption{Resolved category}
\end{subfigure}
\caption{
Distributions of \msmin in the  signal region for the merged~(a) and resolved~(b) category, before any fit~(`Pre-fit').
The contributions from all SM backgrounds are shown as a histogram stack.
`Multiboson' includes contributions from \(VV\), \(VVV\), and electroweak \(VVjj\) production.
The hatched bands represent the total MC statistical uncertainty of the SM expectation.
The expected signal from a representative set of dark Higgs models with \(\gq=0.25\), \(\gchi=1\), \(\mDM=200~\GeV\), and \(\sin\theta=0.01\) scaled for presentation purposes is also shown for reference.
All selection criteria for the corresponding category (see \Tabs{\ref{tab:mergedRegions}}{\ref{tab:resolvedRegions}}) are applied except for the binning in \msmin.
\label{fig:msmin}
}
\end{figure}

An exact kinematic reconstruction of the four-momentum of the dark Higgs particle \(s\) is not possible because three invisible particles in the final state contribute to \ptmiss: the neutrino from the \(W\to\ell\nu\) decay and the two DM particles \(\chi\).
Hence, a dedicated method is used to calculate the minimum possible dark Higgs mass \msmin given the four-momentum of the hadronically decaying \(W\) boson candidate \Wcand, the lepton four-momentum, and the invariant mass of the charged lepton and neutrino being equal to \(m_W\).
Neglecting the lepton mass, the solution for the neutrino energy in this approach is
\begin{equation}
\label{eq:neutrinoEnergy}
E_\nu = \frac{m_W^2}{2E_\ell (1-\cos{\theta_{\ell\nu}})}\,.
\end{equation}
After a rotation into a coordinate system where the charged lepton travels along the \(z\)-axis and the \Wcand momentum is in the \(xz\)-plane, the neutrino four-momentum becomes
\begin{equation*}
p_\nu = \frac{m^2_W}{2 E_\ell (1-\cos{\theta_{\ell\nu}})} (\sin{\theta_{\ell\nu}} \cos{\phi_\nu}, \sin{\theta_{\ell\nu}} \sin{\phi_\nu}, \cos{\theta_{\ell\nu}}, 1)\,.
\end{equation*}
The invariant mass of the \(s\rightarrow WW\) system is then
\begin{equation} \label{eq:ms}
\begin{aligned}
m_s^2 & = && (p_{\Wcand} + p_\ell + p_\nu)^2 \\
& = && (E_{\Wcand} + E_\ell + E_\nu)^2 - (p_{x,\Wcand} + E_\nu \sin{\theta_{\ell\nu}} \cos{\phi_\nu})^2 \\
&  && - (E_\nu \sin{\theta_{\ell\nu}} \sin{\phi_\nu})^2 - (E_\ell + p_{z,\Wcand} + E_\nu \cos{\theta_{\ell\nu}})^2 .
\end{aligned}
\end{equation}
It can be shown that the minimum occurs when \(\phi_\nu = 0\). Using this and \Eqn{(\ref{eq:neutrinoEnergy})}, \Eqn{(\ref{eq:ms})} can be written as
\begin{equation*}
\begin{split}
m_s^2 = \left( E_\ell + \frac{m_W^2}{2E_\ell (1-\cos{\theta_{\ell\nu}})} + E_{\Wcand} \right)^2 - \left( \abs{\vec{p}_{\Wcand}} \sin{\theta_{\Wcand\ell}} + \frac{m^2_{W}\sqrt{1-\cos^2 \theta_{\ell\nu} } }{2E_\ell (1-\cos{\theta_{\ell\nu}}) } \right)^2 \\
- \left( E_\ell + \abs{\vec{p}_{\Wcand}} \cos{\theta_{\Wcand\ell}} + \frac{m^2_W \cos{\theta_{\ell\nu}} }{2E_\ell(1-\cos{\theta_{\ell\nu}}) } \right)^2 ,
\end{split}
\end{equation*}
which leaves only \(\cos{\theta_{\ell\nu}}\) as an unknown. Minimizing this function over \(\cos{\theta_{\ell\nu}}\) leads therefore to the desired observable \msmin:
\begin{equation*}
\msmin \equiv \min_{\cos{\theta_{\ell\nu}}} (\ms) .
\end{equation*}
The distributions of \msmin in the merged and resolved categories of the SR are shown in \Fig{\ref{fig:msmin}}.

The \msmin observable is used as the final discriminant in the SR to separate the signal, which is resonant in \msmin, from the backgrounds, which tend to be non-resonant in \msmin.
To effectively use the \msmin information, both the merged and resolved categories of the SR are split into bins.
Taking into account the expected number of events and their predicted distribution in \msmin, the merged category is split into four  \msmin bins of unequal width, as indicated in \Tab{\ref{tab:mergedRegions}}.
In a similar spirit, the resolved category is divided into five equal-width bins of \SI{50}{\GeV} in \msmin, as displayed in \Tab{\ref{tab:resolvedRegions}}.

\subsection{Background estimation}
Control regions are used to improve the modelling of dominant backgrounds. The CRs are split into merged and resolved categories in analogy with the SR to mimic its kinematic properties.
In order to obtain a sufficient number of events in the CRs for the merged category, the requirement on \metsig is lowered to \(\mathcal{S} > 12\), while all other selection requirements remain identical to those for the SR.
It was confirmed that with this change the events in the CRs still closely mimic the kinematic properties of the SRs.

In order to improve the MC prediction for the dominant \wjets background, a dedicated control region is used to normalize this process.
This CR with suppressed signal contribution is defined by requiring \(\dRTARl>1.8\) in the merged category and \(\dRWl>1.4\) in the resolved category.
In a similar fashion, a CR is used for the sub-dominant \ttbar process, and requires at least two \(b\)-tagged jets.
The exact definitions of the CRs can be found in \Tabs{\ref{tab:mergedRegions}}{\ref{tab:resolvedRegions}}.
Other sub-dominant and minor backgrounds include \zjets, diboson, triboson and single-top processes.
The predictions for these backgrounds are based purely on MC simulations, normalized to their respective theoretical cross sections.
About 20\% of the total background is contributed by multiboson production, of which about 90\% is diboson production.
Generally, good agreement between data and MC simulation is found in the CRs, given the uncertainties.
A potential contribution from QCD multijet production was studied and found negligible.

To predict the SM background yields in the SR, a simultaneous profile likelihood fit~\cite{Cowan:2010js} is performed to constrain the MC yields with the observed data in the CRs, using standard minimization algorithms~\cite{RooFit,2015baak}.
Four background normalization factors are used separately for the \wjets and \ttbar processes in the merged and resolved regimes, and are allowed to float freely in the fit.
The systematic uncertainties discussed in Section~\ref{sec:systematics} are parameterized as nuisance parameters with Gaussian or log-normal prior probabilities in the fit.
%
%
%
%

% End of text imported from the .//sections/analysis.tex input file

%
\section{Systematic uncertainties}
\label{sec:systematics}

% The next lines are included from the .//sections/systematics.tex input file
In the following, the estimation of the relevant systematic uncertainties for background and signal processes is discussed.
The systematic effects are studied coherently across all analysis regions.
The impact of the uncertainties is quantified as the fractional effect relative to the total background yield in the \merged (\resolved) category of the SR before any fit, using an X\%~(Y\%) format.

A major source of uncertainty is the statistical uncertainty of the MC simulation, which amounts to 9\%~(5\%). This is due to the very low acceptance of the analysis to dominant SM backgrounds like \wjets, which is the main contributor to this uncertainty category.
Theoretical uncertainties of the \wjets and \ttbar backgrounds that affect their normalization are significantly reduced through the use of CRs and are limited by the statistical uncertainty of the data and the extrapolation to the SR.
The evaluation of theoretical modelling uncertainties is described in \Sect{\ref{sec:data-mc}}.
The leading theoretical uncertainties are the uncertainties related to the matrix element generator for the \ttbar sample, amounting to 4\% (1\%), as well as for the single-top sample, amounting to 3\% (1\%), the diboson scale uncertainty, amounting to 3\% (2\%), and the \wjets scale uncertainty, contributing 2\% (6\%).
The leading experimental systematic uncertainties are related to the jet energy scale (JES) and  jet energy resolution (JER).
These uncertainties are derived as a function of the \pt and \(\eta\) of the jet, the \pileup conditions, and the jet-flavour composition of the selected sample.
They are estimated %
separately for \(R=0.4\) PFlow jets~\cite{JETM-2018-05} and \(R=0.2\) LCW jets, following methodologies similar to those in \Refn{\cite{ATL-PHYS-PUB-2015-015}}.
The JES and JER uncertainties for R=0.2 jets are propagated to the observables related to TAR jets like \Wcandm, \msmin and \Dtwo.
The effect of the JER uncertainty amounts to 9\% (5\%), and that of the JES uncertainty to 7\% (5\%).
Sub-dominant experimental uncertainties considered are associated with the modelling of \met~\cite{PERF-2016-07}, amounting to 3\% (1\%), the track reconstruction efficiency~\cite{PERF-2015-08} (3\% in the \merged category) and the luminosity uncertainty~\cite{ATLAS-CONF-2019-021} (2\% in the both categories).
Other considered, but less important, uncertainties are associated with the modelling of \(b\)-tagging efficiencies~\cite{PERF-2016-05}, the reconstruction efficiency, energy scale, energy resolution, and trigger efficiency of leptons~\cite{EGAM-2018-01,MUON-2018-03,TRIG-2016-01,TRIG-2018-01}, the \pileup reweighting, and the tagging and suppression of \pileup jets~\cite{ATLAS-CONF-2014-018}.

While quantifying the impact of a given systematic uncertainty as the fractional effect relative to the total background yield in each SR category is useful, it does not capture effects that are differential in \msmin.
Hence, the effect of systematic uncertainties is quantified in \Tab{\ref{tab:syst-breakdown}} in terms of their contribution to the fitted signal uncertainty relative to the theory prediction.
In this procedure, each systematic uncertainty's squared contribution is given by the difference between the squares of the total uncertainty and the uncertainty obtained by neglecting the systematic uncertainty source in question, where in the latter fit the signal value is fixed to its value considering all uncertainties.
This is done for three representative signal models that span the range of the dark Higgs model parameter space that this analysis is sensitive to.

\begin{table}[htbp]
\centering
\caption{
Dominant sources of uncertainty for three dark Higgs scenarios after the fit to data.
The uncertainties are quantified in terms of their contribution to the fitted signal uncertainty, which is expressed relative to the theory prediction.
Three representative dark Higgs signal scenarios with \(\gq=0.25\), \(\gchi=1.0\), \(\sin\theta=0.01\) and \(\mDM=\SI{200}{GeV}\) are considered, which are indicated using the \((\mZp,\,\ms)\) format in units of GeV in the table columns.
`Total uncertainty' is the quadrature sum of statistical and total systematic uncertainties, which consider correlations.
Only the largest systematic uncertainties are shown.
}
\label{tab:syst-breakdown}
\begin{tabular}{l c c c}
\toprule
\multirow{2}{*}{Source of uncertainty} & \multicolumn{3}{c}{Uncertainty [\%]} \\
& \((2100,\,210)\)  & \((1000,\,140)\) & \((1000,\,360)\) \\
\midrule
\wjets modelling & 4 & 5 & 2 \\
Diboson modelling & 5 & 4 & 1 \\
\ttbar modelling & 7 & 4 & 1 \\
Single top modelling & 9 & 5 & 11~~ \\
Signal modelling & 1 & 3 & 0 \\
Statistical uncertainty of MC & 26~~ & 15~~ & 29~~ \\
\(R=0.4\) jet energy scale & 11~~ & 12~~ & 14~~ \\
\(R=0.4\) jet energy resolution & 9 & 4 & 7 \\
\(R=0.2\) jet energy scale & 9 & 9 & 14~~ \\
\(R=0.2\) jet energy resolution & 13~~ & 10~~ & 16~~ \\
\met & 7 & 1 & 7 \\
Track reconstruction & 5 & 2 & 2 \\
Lepton reconstruction & 2 & 3 & 1 \\
\midrule
Systematic uncertainty & 38~~ & 28~~ & 40~~ \\
Statistical uncertainty of data & 38~~ & 32~~ & 37~~ \\
Total uncertainty & 53~~ & 43~~ & 55~~ \\
\bottomrule
\end{tabular}
\end{table}

% End of text imported from the .//sections/systematics.tex input file

%
\section{Results}
\label{sec:results}

\begin{figure}
\centering
\includegraphics[width=0.7\textwidth]{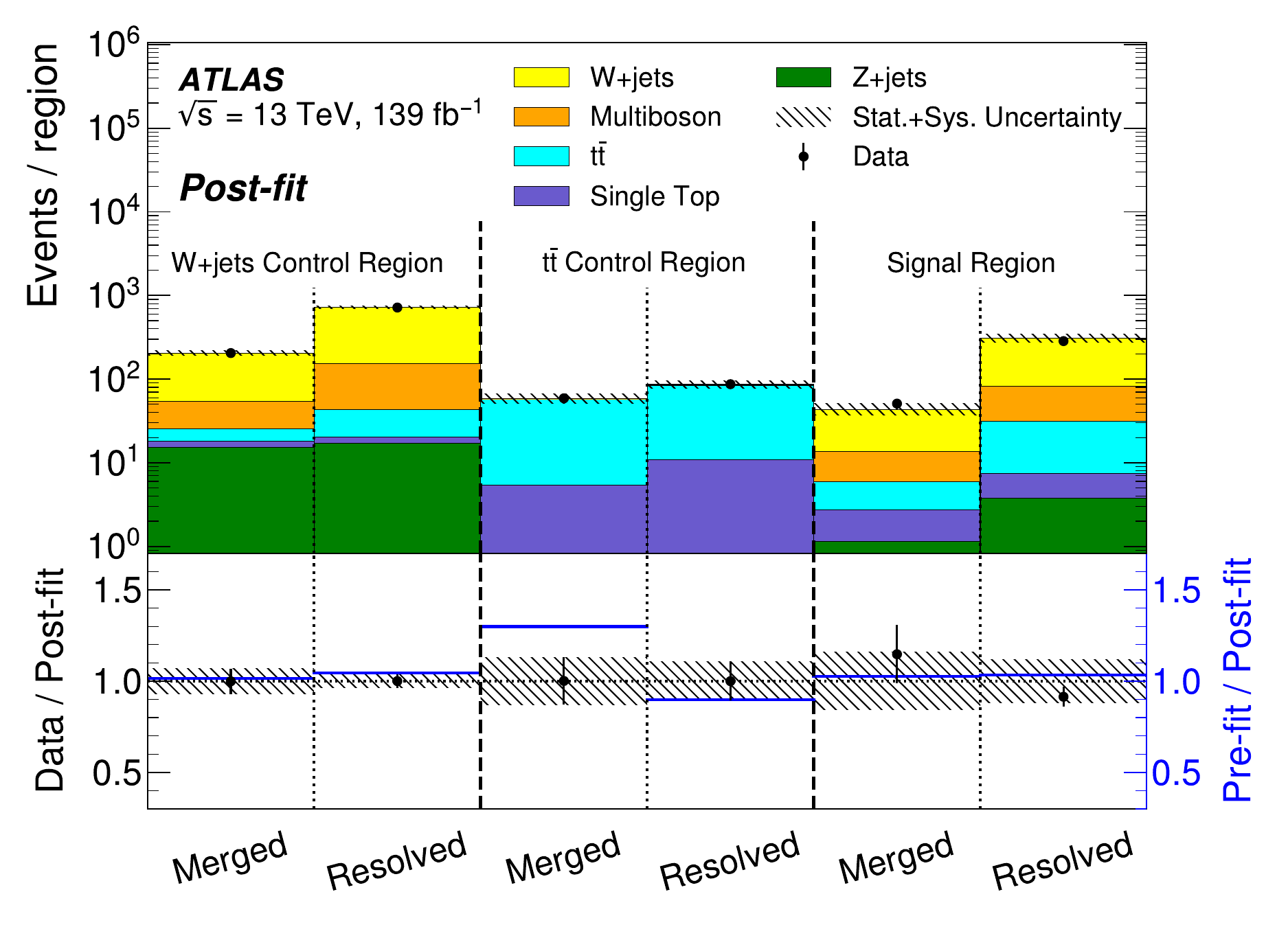}
\caption{
Data overlaid on SM background yields stacked in each SR and CR category after the fit to data~(`Post-fit'). The maximum-likelihood estimators are set to the conditional values of the CR-only fit, and propagated to the SR and CRs.
The lower panel displays the ratio of data to SM expectations after the fit, with its systematic uncertainty
considering correlations between individual contributions
indicated by the hatched band.
The lower panel also displays the blue line representing the ratio of pre-fit to post-fit background predictions.
}
\label{fig:overview_cr_sr}
\end{figure}

\begin{figure}
\centering
\begin{subfigure}{0.48\textwidth}
\includegraphics[width = \textwidth]{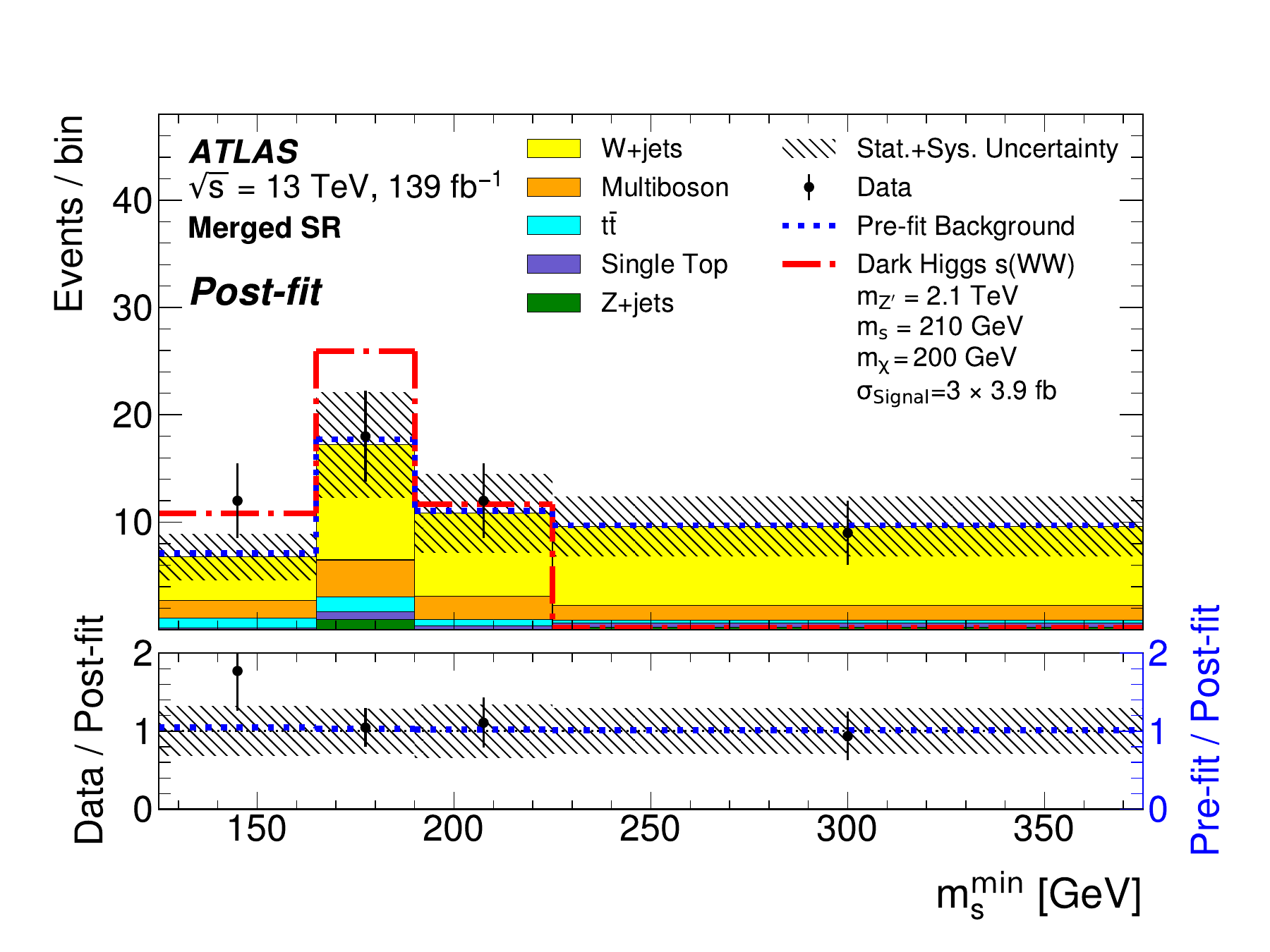}
\caption{Merged category}
\end{subfigure}
\quad
\begin{subfigure}{0.48\textwidth}
\includegraphics[width = \textwidth]{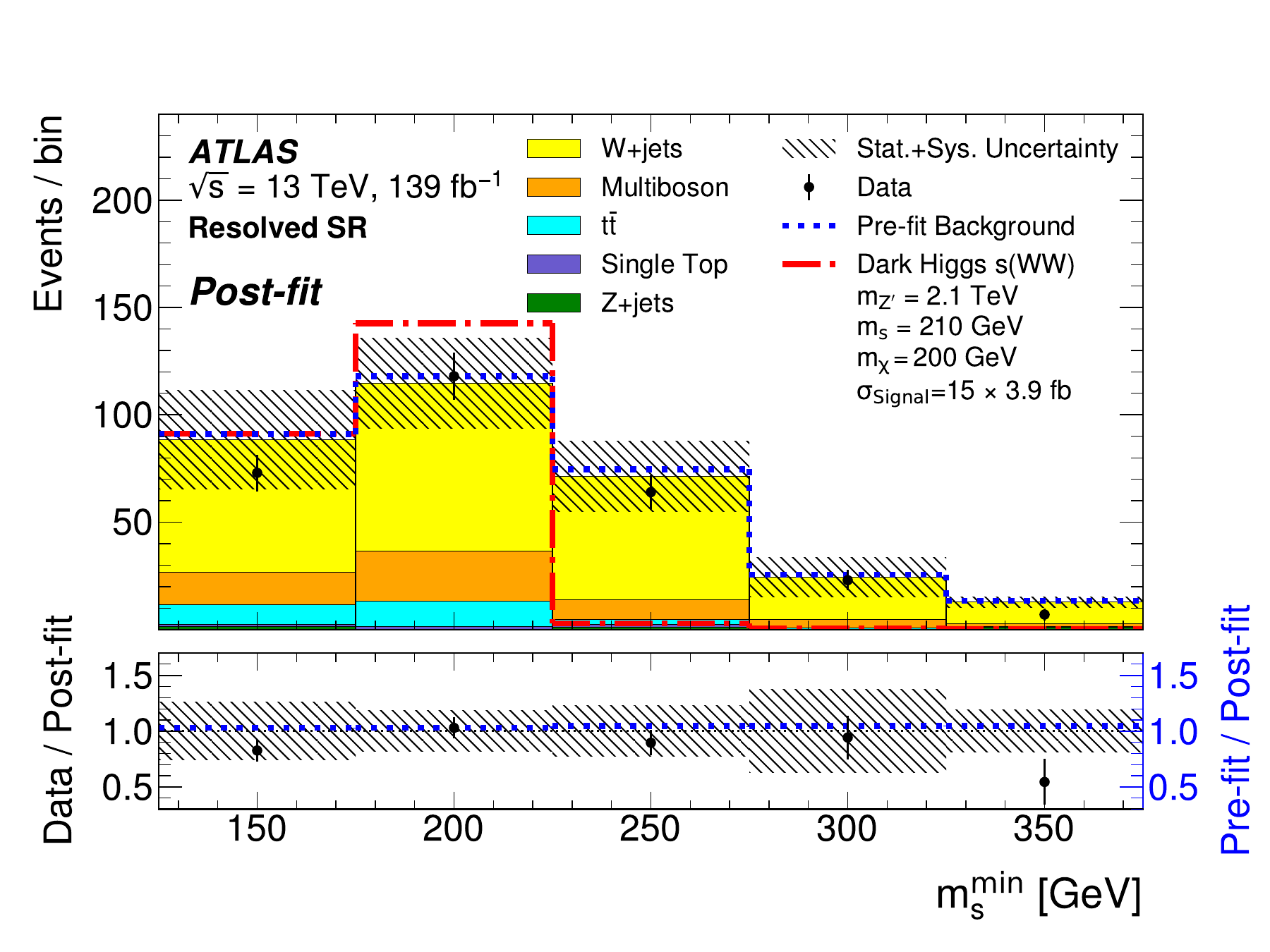}
\caption{Resolved category}
\end{subfigure}
\caption{
Distributions of the invariant mass of the dark Higgs candidates in the signal region for the merged~(a) and resolved~(b) category, after the fit to data~(`Post-fit'). The upper panels compare the data with the SM expectation before (blue dashed line) and after the background-only fit (histogram stack). The lower panels display the ratio of data to SM expectations after the fit, with its systematic uncertainty. Also shown is the ratio of SM expectations before and after the fit~(blue dashed line).
The expected signal from a representative dark Higgs model with \(\gq=0.25\), \(\gchi=1\), \(\mDM=200~\GeV\), and \(\sin\theta=0.01\) assuming \(\mZp=2.1\)~TeV and \(m_s=210~\GeV\), with a cross section of 3.9~fb for the  \(s\to W^-(\ell^-\bar\nu) W^+(\qpqbar)\) decay mode, is shown as the dash-dotted line and scaled for presentation purposes.
}
\label{fig:sr_after}
\end{figure}

% The next lines are included from the .//sections/results.tex input file
The data analysis is performed in several steps and with the SR initially blinded to minimize any human bias.
First, a fit to the SM backgrounds is performed only with data from the \wjets and \ttbar CRs, split into the merged and resolved categories analogously to the SR.
The resulting observed and fitted yields in the CR categories are shown in Figure~\ref{fig:overview_cr_sr}.
To demonstrate the propagation of constraints on the \wjets and \ttbar backgrounds and experimental uncertainties from the CRs to the SR, Figure~\ref{fig:overview_cr_sr} also shows the yields of background processes predicted in the SR when using the observed parameter values from the CR-only fit.
This fit slightly reduces the overall \vjets contribution.
The fitted normalization of the dominant \wjets background is \(0.99\pm 0.17\) \((0.94\pm 0.07)\) in the merged (resolved) category.
The \ttbar background normalization falls to \(0.75\pm0.13\) in the merged category, and rises to \(1.14\pm 0.22\) in the resolved category.
Given the uncertainties, no significant discrepancy between the yields in the SR categories is observed.
Finally, a simultaneous fit to the SR and the CRs is performed considering both the merged and resolved categories.
The reconstructed dark Higgs candidate mass \msmin is used as the main discriminant between signal and background in the SR in this fit set-up.
Figure~\ref{fig:sr_after} shows the \msmin distributions of the dark Higgs candidate in the merged and resolved SR categories, obtained after a simultaneous fit to the SR and the CRs under the hypothesis that only the SM predictions are present.
In the combined SR and CR fit the data distributions are found to be well described by MC simulations within the estimated uncertainties, similarly to the CR-only fit.
The data are found to be one standard deviation above the MC simulations in the merged category for \(\msmin<165~\GeV\), but this trend is not reproduced in the resolved category.
A data event deficit of about 1.5 standard deviations is observed for \(\msmin>325~\GeV\) in the resolved category.

\begin{figure}
\centering
\includegraphics[width=0.6\textwidth]{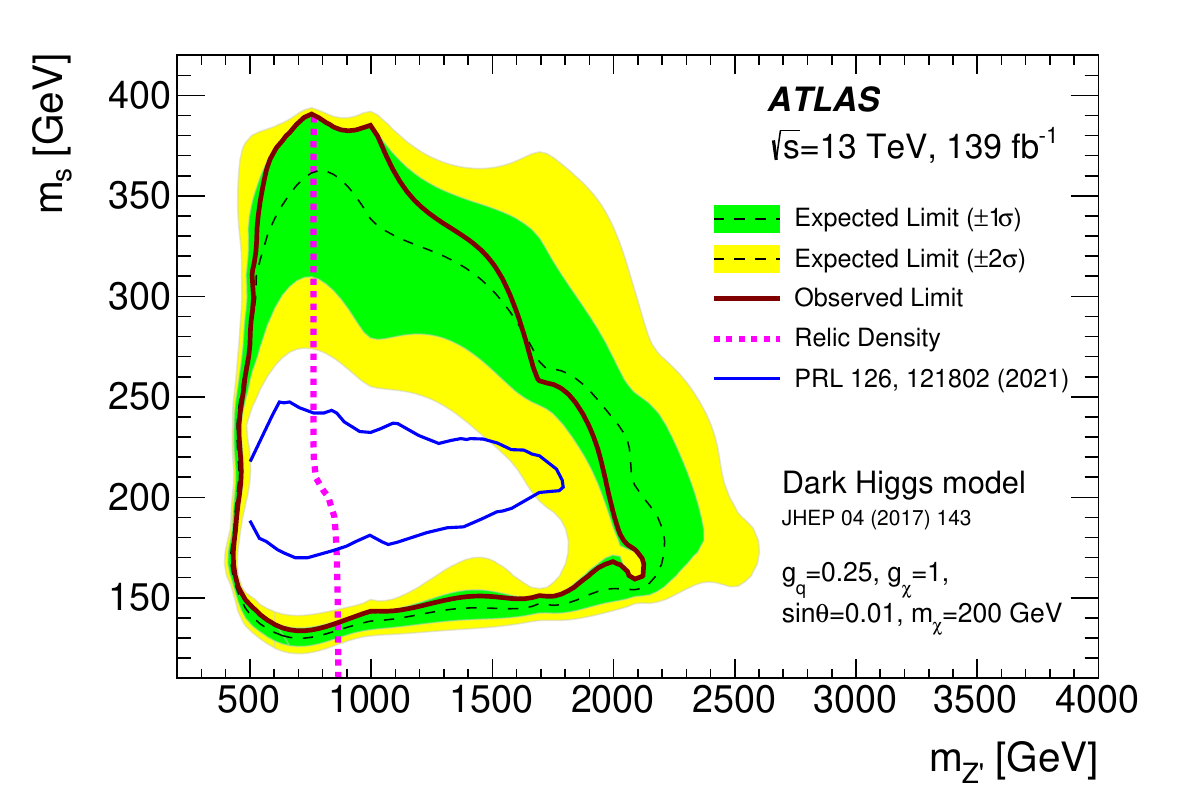}
\caption{
Exclusion contours for the dark Higgs model in the \((\mZp,\,\ms)\) plane for \(\gq=0.25\), \(\gchi=1\), \(\mDM=200~\GeV\), and \(\sin\theta=0.01\).
The observed (expected) \(95\%\) CL exclusion is represented by the solid line (dashed line), along with the \(\pm1\sigma\) (\(\pm2\sigma\)) expected uncertainty as the filled green (yellow) band.
The area within the contour curve is excluded.
The parameter values where the relic density value, as calculated with MadDM~\cite{Arina:2021gfn}, matches the observed relic density~\cite{Aghanim:2018eyx}  are shown by the short-dashed line.
Values of \mZp above this line correspond to an overabundance of DM.
The observed exclusion contour from the analysis of the \(s\to VV\to q\bar q'\! q''\!\bar q'''\) decay channel~\cite{EXOT-2018-40}, where \(VV=W^+W^-\) or \(ZZ\), is also shown.
}
\label{fig:limit}
\end{figure}

The results of this search are interpreted by setting upper limits on the product of the \(pp\rightarrow s \chi\chi\) production cross section and the decay branching fraction \(\mathcal{B}(s\to W^+W^-)\), using a modified frequentist approach~(CL\(_\mathrm{s}\))~\cite{Read:2002hq} with a test statistic based on the profile likelihood in the asymptotic approximation~\cite{Cowan:2010js}.
Upper limits on the ratio of the measured signal cross section to its theoretically predicted value are determined at \(95\%\) confidence level (CL).
Exclusion contours in the (\mZp, \ms) plane for the dark Higgs model with \(\gq=0.25\), \(\gchi=1.0\), and \(\sin\theta=0.01\)
are presented in Figure~\ref{fig:limit}.
Consistency with the observed relic density can indicate the preferred model parameters if it is assumed that no additional particles beyond $s$ and $\Zp$ are present.
For $\ms \lesssim 200~\GeV$, the relic density suggests \mZp masses of about $850~\GeV$, and increases with growing \mZp. For $\ms \gtrsim 200~\GeV$, a lower \mZp value of about 760~\GeV\ is preferred for consistency with the observed relic density; this is due to annihilation of WIMP pairs into pairs of on-shell dark Higgs bosons becoming kinematically impossible.
The sensitivity of the search is highest for \(\mZp\approx750\)~\GeV\ and for this \mZp value a wide range of dark Higgs boson masses from 140~\GeV\ to 390~\GeV\ can be excluded.
The existence of a hypothetical \Zp boson with a mass of up to \(1.8~\TeV\) is excluded for \(150\lesssim\ms\lesssim250\)~\GeV\ at \(95\%\) CL.
In this dark Higgs boson mass range, the observed exclusion at high \mZp is weaker than expected owing to the small excess in data at \(\msmin<165~\GeV\) in the merged region that was discussed above.
For $\ms \gtrsim 300~\GeV$, the exclusion limits are somewhat stronger than expected owing to the deficit in data  at \(\msmin>325~\GeV\) in the resolved region.

% End of text imported from the .//sections/results.tex input file

%

%
%
\FloatBarrier

\section{Conclusion}
\label{sec:conclusion}

% The next lines are included from the .//sections/conclusion.tex input file
A search for dark matter in final states with large \met and semileptonic decays of resonantly produced \(W^\pm W^\mp\) pairs is presented.
The search uses the full Run~2 data set recorded with the ATLAS detector at the LHC in 2015--2018, which corresponds to \SI{139}{\ifb} of \(pp\) collisions at \(\sqrt{s} = \SI{13}{TeV}\).
A new technique of track-assisted reclustering was used to identify merged hadronic \(W\) boson candidates as a single large-radius jet, in addition to a resolved method using a combination of small-radius jets.
The data are found to be consistent with the Standard Model predictions.
The result is interpreted as upper limits at 95\% confidence level on the dark Higgs model parameters.
The most stringent limits on \(m_{Z'}\) can be set for \(m_s = \SI{160}{GeV}\), excluding the range \(500 < \mZp < 2100\)~\GeV, and the most stringent limits on \(m_s\) can be set for \(\mZp = \SI{750}{GeV}\), excluding the range \(140 < \ms < 390\)~\GeV.
This result substantially extends the sensitivity into mass regions beyond the reach of previous analyses by the ATLAS and CMS collaborations in fully hadronic final states and fully leptonic final states for dark Higgs boson masses above 140~\GeV.

% End of text imported from the .//sections/conclusion.tex input file

%

%
\section*{Acknowledgements}
%

% The next lines are included from the .//acknowledgements/Acknowledgements.tex input file
%
%

%
%

We thank CERN for the very successful operation of the LHC, as well as the
support staff from our institutions without whom ATLAS could not be
operated efficiently.

We acknowledge the support of
ANPCyT, Argentina;
YerPhI, Armenia;
ARC, Australia;
BMWFW and FWF, Austria;
ANAS, Azerbaijan;
CNPq and FAPESP, Brazil;
NSERC, NRC and CFI, Canada;
CERN;
ANID, Chile;
CAS, MOST and NSFC, China;
Minciencias, Colombia;
MEYS CR, Czech Republic;
DNRF and DNSRC, Denmark;
IN2P3-CNRS and CEA-DRF/IRFU, France;
SRNSFG, Georgia;
BMBF, HGF and MPG, Germany;
GSRI, Greece;
RGC and Hong Kong SAR, China;
ISF and Benoziyo Center, Israel;
INFN, Italy;
MEXT and JSPS, Japan;
CNRST, Morocco;
NWO, Netherlands;
RCN, Norway;
MEiN, Poland;
FCT, Portugal;
MNE/IFA, Romania;
MESTD, Serbia;
MSSR, Slovakia;
ARRS and MIZ\v{S}, Slovenia;
DSI/NRF, South Africa;
MICINN, Spain;
SRC and Wallenberg Foundation, Sweden;
SERI, SNSF and Cantons of Bern and Geneva, Switzerland;
MOST, Taiwan;
TENMAK, T\"urkiye;
STFC, United Kingdom;
DOE and NSF, United States of America.
In addition, individual groups and members have received support from
BCKDF, CANARIE, Compute Canada and CRC, Canada;
PRIMUS 21/SCI/017 and UNCE SCI/013, Czech Republic;
COST, ERC, ERDF, Horizon 2020 and Marie Sk{\l}odowska-Curie Actions, European Union;
Investissements d'Avenir Labex, Investissements d'Avenir Idex and ANR, France;
DFG and AvH Foundation, Germany;
Herakleitos, Thales and Aristeia programmes co-financed by EU-ESF and the Greek NSRF, Greece;
BSF-NSF and MINERVA, Israel;
Norwegian Financial Mechanism 2014-2021, Norway;
NCN and NAWA, Poland;
La Caixa Banking Foundation, CERCA Programme Generalitat de Catalunya and PROMETEO and GenT Programmes Generalitat Valenciana, Spain;
G\"{o}ran Gustafssons Stiftelse, Sweden;
The Royal Society and Leverhulme Trust, United Kingdom.

The crucial computing support from all WLCG partners is acknowledged gratefully, in particular from CERN, the ATLAS Tier-1 facilities at TRIUMF (Canada), NDGF (Denmark, Norway, Sweden), CC-IN2P3 (France), KIT/GridKA (Germany), INFN-CNAF (Italy), NL-T1 (Netherlands), PIC (Spain), ASGC (Taiwan), RAL (UK) and BNL (USA), the Tier-2 facilities worldwide and large non-WLCG resource providers. Major contributors of computing resources are listed in Ref.~\cite{ATL-SOFT-PUB-2021-003}.

%
%

% End of text imported from the .//acknowledgements/Acknowledgements.tex input file

%
%
%
%
%
%
%

%
\clearpage
\printbibliography
\clearpage
% ATLAS Collaboration author list
% Reference date of EXOT-2020-04 is 2022-04-22
% Author list last updated on date 06-JUL-23
% Data extracted on 06-Jul-2023 for paper reference EXOT-2020-04
% at 1:59pm
 
\begin{flushleft}
\hypersetup{urlcolor=black}
{\Large The ATLAS Collaboration}

\bigskip

\AtlasOrcid[0000-0002-6665-4934]{G.~Aad}$^\textrm{\scriptsize 101}$,
\AtlasOrcid[0000-0002-5888-2734]{B.~Abbott}$^\textrm{\scriptsize 119}$,
\AtlasOrcid[0000-0002-7248-3203]{D.C.~Abbott}$^\textrm{\scriptsize 102}$,
\AtlasOrcid[0000-0002-1002-1652]{K.~Abeling}$^\textrm{\scriptsize 55}$,
\AtlasOrcid[0000-0002-8496-9294]{S.H.~Abidi}$^\textrm{\scriptsize 29}$,
\AtlasOrcid[0000-0002-9987-2292]{A.~Aboulhorma}$^\textrm{\scriptsize 35e}$,
\AtlasOrcid[0000-0001-5329-6640]{H.~Abramowicz}$^\textrm{\scriptsize 150}$,
\AtlasOrcid[0000-0002-1599-2896]{H.~Abreu}$^\textrm{\scriptsize 149}$,
\AtlasOrcid[0000-0003-0403-3697]{Y.~Abulaiti}$^\textrm{\scriptsize 116}$,
\AtlasOrcid[0000-0003-0762-7204]{A.C.~Abusleme~Hoffman}$^\textrm{\scriptsize 136a}$
% ...
\bigskip
\\

% ...
$^{29}$Physics Department, Brookhaven National Laboratory, Upton NY; United States of America.\\
% ...
$^{35}$$^{(a)}$Facult\'e des Sciences Ain Chock, R\'eseau Universitaire de Physique des Hautes Energies - Universit\'e Hassan II, Casablanca;$^{(b)}$Facult\'{e} des Sciences, Universit\'{e} Ibn-Tofail, K\'{e}nitra;$^{(c)}$Facult\'e des Sciences Semlalia, Universit\'e Cadi Ayyad, LPHEA-Marrakech;$^{(d)}$LPMR, Facult\'e des Sciences, Universit\'e Mohamed Premier, Oujda;$^{(e)}$Facult\'e des sciences, Universit\'e Mohammed V, Rabat;$^{(f)}$Institute of Applied Physics, Mohammed VI Polytechnic University, Ben Guerir; Morocco.\\
% ...
$^{55}$II. Physikalisches Institut, Georg-August-Universit\"{a}t G\"ottingen, G\"ottingen; Germany.\\
% ...
$^{101}$CPPM, Aix-Marseille Universit\'e, CNRS/IN2P3, Marseille; France.\\
$^{102}$Department of Physics, University of Massachusetts, Amherst MA; United States of America.\\
% ...
$^{116}$Department of Physics, New York University, New York NY; United States of America.\\
% ...
$^{119}$Homer L. Dodge Department of Physics and Astronomy, University of Oklahoma, Norman OK; United States of America.\\
% ...
$^{136}$$^{(a)}$Departamento de F\'isica, Pontificia Universidad Cat\'olica de Chile, Santiago;$^{(b)}$Millennium Institute for Subatomic physics at high energy frontier (SAPHIR), Santiago;$^{(c)}$Instituto de Investigaci\'on Multidisciplinario en Ciencia y Tecnolog\'ia, y Departamento de F\'isica, Universidad de La Serena;$^{(d)}$Universidad Andres Bello, Department of Physics, Santiago;$^{(e)}$Instituto de Alta Investigaci\'on, Universidad de Tarapac\'a, Arica;$^{(f)}$Departamento de F\'isica, Universidad T\'ecnica Federico Santa Mar\'ia, Valpara\'iso; Chile.\\
% ...
$^{149}$Department of Physics, Technion, Israel Institute of Technology, Haifa; Israel.\\
$^{150}$Raymond and Beverly Sackler School of Physics and Astronomy, Tel Aviv University, Tel Aviv; Israel.\\
% ...
$^{a}$ Also Affiliated with an institute covered by a cooperation agreement with CERN.\\
$^{b}$ Also at An-Najah National University, Nablus; Palestine.\\
$^{c}$ Also at Borough of Manhattan Community College, City University of New York, New York NY; United States of America.\\
$^{d}$ Also at Bruno Kessler Foundation, Trento; Italy.\\
$^{e}$ Also at Center for High Energy Physics, Peking University; China.\\
$^{f}$ Also at Centro Studi e Ricerche Enrico Fermi; Italy.\\
% ...

\end{flushleft}

% Created with Glance <Atlas.Glance@cern.ch>

%

%
%

%

%
%
%
%

\end{document}